\documentstyle[12pt]{article}
\def\lromn#1{\uppercase\expandafter{\romannumeral#1}}

\makeatletter

\renewcommand{\theequation}%
{\arabic{section}.\arabic{equation}}
\@addtoreset{equation}{section}
\renewcommand{\appendix}{\par
  \setcounter{section}{0}
  \setcounter{subsection}{0}
  \renewcommand{\thesection}{Appendix~\Alph{section}}
  \renewcommand{\theequation}{\Alph{section}.\arabic{equation}}}

\makeatother

\begin{document}
\begin{flushright}
TU/98/535\\
RCNS-98-02\\
\end{flushright}

\begin{center}
\begin{large}

\bf{
Prolonged Decay and CP-asymmetry
}

\end{large}

\vspace{36pt}

\begin{large}
I. Joichi, Sh. Matsumoto, and M. Yoshimura

Department of Physics, Tohoku University\\
Sendai 980-8578 Japan\\
\end{large}

\vspace{54pt}

{\bf ABSTRACT}

\end{center}

Time evolution of unstable particles that occur in the
expanding universe is investigated.
The off-shell effect not included in the Boltzmann-like
equation is important for the decay process when the temperature
becomes much below the mass of unstable particle.
When the off-shell effect is taken into account,
the thermal abundance of unstable particles at low temperatures
has a power law behavior of temperature $T$, 
$\frac{\Gamma }{M}(\frac{T}{M})^{\alpha + 1}$
unlike the Boltzmann suppressed $e^{-M/T}$,
with the power $\alpha $ related to the spectral rise near the
threshold of the decay and with $\Gamma $ the decay rate.
Moreover, the relaxation time towards the thermal value
is not governed by the exponential law; instead, 
it is the power law of time.
The evolution equation for the occupation number and the number
density of the unstable particle is derived, when both of these effects, 
along with the cosmic expansion, are included. 
We also critically examine how the scattering off thermal particles
may affect the off-shell effect to the unstable particle.
As an application showing the importance of the off-shell effect
we compute the time evolution of the baryon asymmetry generated by
the heavy $X$ boson decay.
It is shown that the out-of equilibrium kinematics previously discussed
is considerably changed.


\newpage
\section{Introduction}

\vspace{0.5cm} 
\hspace*{0.5cm} 
There are many short-lived particles that have existed in abundance
in the early universe whose temporary presence did not leave behind
any measurable effect.
Important exceptions to this exist, such as 
the neutron which certainly is the key for explanation of
element abundance of the present universe.

Theoretical estimate of the abundance  of these unstable particles 
after the cosmic temperature drops below the mass of the unstable
particle is very important
for subsequent time evolution. Most works in the past \cite{kolb-turner book}
are based on the Boltzmann equation that takes into account relevant
reactions in the expanding universe.
The use of the Boltzmann equation has however been questioned recently
\cite{jmy-96};
a more precise quantum mechanical description of the decay process
in thermal medium should contain important off-shell contribution not
properly treated in the Boltzmann approach.
These off-shell effects are eminent in the low temperature region.
Low temperature effects are clearly important in this problem, since
unstable particles  are typically very non-relativistic
when they disappear in the early universe.

In the present work we shall develop a general formalism of computing
time evolution of the net number density of unstable particles
and clarify the off-shell effect.
The off-shell effect appears in two ways; first, in a slower relaxation
towards the equilibrium abundance and second, in  a larger equilibrium
value not suppressed by the Boltzmann factor such as
\( \:
e^{-\,\Delta M/T}
\: \)
where $\Delta M$ is the mass difference of the parent and the daughter
particles.
It is shown below that the off-shell effect becomes dominant 
below some temperature $T_{{\rm eq}}$.
The abundance of unstable particles then follows the power law;
\( \:
\frac{n}{T^{3}} \approx  \frac{\Gamma }{M}(\frac{T}{M})^{\alpha + 1}
\,, 
\: \)
where $\alpha $ is a parameter related to the threshold behavior of
the spectral function for the decay and $\Gamma $ is the decay rate.
Thus, unstable particles do not disappear suddenly. Instead, their
abundance gradually decreases with a power of decreasing temperature
as the universe expands.
Physical processes that follow after the decay are then prolonged.
The off-shell effect turns out to be more prominent for a larger decay rate.

We next consider as an illustrative application of this general result
the hypothetical $X$ boson decay
that may have created the matter-antimatter asymmetry when they decay
\cite{gut b-asymmetry}, \cite{b-asymmetry review}.
We find that the time evolution of the baryon asymmetry is
substantially changed and the severe lower bound of the $X$ boson mass is 
considerably relaxed by the off-shell effect.
For the first time we find that some mode of the X boson decay
for baryogenesis is excluded due to the off-shell effect.
This is the S-wave decay mode into a boson-pair.

This paper is organized as follows.
In Section 2 the theoretical model of unstable particle decay
is explained. This is a field theoretical extention of the harmonic model
for the quantum dissipation in thermal medium 
discussed in \cite{jmy-96}. We first present and formally solve the 
quantum mechanical model of the decay of excited levels in thermal medium.
A great virture of this model is that its integrability
leads to explicit formulas for many quantities of interest.
One can clearly see how the off-shell effect arises in these formulas.
Extention to the unstable particle decay in field theory models
can be made, but it is in general complicated 
and not readily solvable. But fortunately, in a thermal
medium far away from the degeneracy limit which is relevant in the early
universe the decay process is approximately described by this class of solvable
quantum mechanical models extended to infinitely many decay channels.
In Section 3 the occupation number and the number density of a species
of unstable particles is calculated and its time evolution equation
is derived in the expanding universe. The stationary abundance
when the cosmic expansion is switched off is worked out, and
its behavior at both high and low temperatures is studied in detail.
In Section 4 we pay a special attention to the off-shell effect and
its role in cosmology.
We also discuss a possible effect of the incoherence 
due to the scattering off
thermal particles and its role to the decay process in thermal medium.
In Section 5 we apply previous results to the problem of baryogenesis.
The time evolution equation for the baryon asymmetry is derived,
including the off-shell effect.
This equation is analyzed both analytically and numerically, and
a comparison is made when only the on-shell contribution is retained.

\vspace{0.5cm} 
\section{Model of unstable particle decay}
\vspace{0.5cm} 
\hspace*{0.5cm} 
We first present an exactly solvable model of the decay of excited levels
in thermal medium, an extention being made to the case of many decay channels.
This is a slight extention of our previous harmonic model \cite{jmy-96}.
We then explain how the two-particle decay of unstable particles in
any quantum field theory can be approximately described by this class of
quantum mechanical models of infinitely many decay channels.
The approximation is valid for the thermal medium of the low occupation
number, the circumstance far away from the degeneracy limit.

We assume that an excited state $|1\rangle $
of energy $E_{1}$ is given by applying a creation operator to the 
vacuum $|0\rangle $;
\( \:
|1\rangle = c^{\dag }|0\rangle \,.
\: \)
There are continuously many states degenerate with this,
\( \:
b^{\dag }(\omega )|0\rangle 
\: \)
as decay states. Here $\omega $ is the energy of the continuous many levels.
Without specifying the nature of the decay process, one may take
the Hamiltonian that governs the decay as
\begin{equation}
H = E_{1}\,c^{\dag }c + \int_{\omega _{c}}^{\infty }\,d\omega \,
\omega \,
b^{\dag }(\omega )b(\omega ) + \int_{\omega _{c}}^{\infty }\,d\omega \,
\sqrt{\sigma (\omega )}\,\left( \,b^{\dag }(\omega )c + c^{\dag }b(\omega )
\,\right) \,.
\end{equation}
Here $\omega _{c}$ is the threshold of the continuous states taken to be
\( \:
\omega _{c} < E_{1} \,,
\: \)
and $\sigma (\omega )$ characterizes the decay interaction.
This Hamiltonian is a general one with regard to the decay process,
complications being hidden in identification of the composite
operator $b^{\dag }(\omega )$ and the spectral
form of interaction $\sigma (\omega )$.

As emphasized elsewhere \cite{jmy-97-1}, 
the dynamical system thus specified is
exactly solvable; one may explicitly construct the diagonal operator
$B^{\dag }(\omega )$ and the eigenstate 
\( \:
|\omega \rangle _{S} = B^{\dag }(\omega )|0\rangle 
\: \)
that diagonalizes the decay Hamiltonian;
\begin{eqnarray}
&& \hspace*{-1.5cm}
B^{\dag }(\omega ) = b^{\dag }(\omega ) +
F(\omega + i0^{+})\,\left( \,-\,\sqrt{\sigma (\omega )}\,c^{\dag }
+ \int_{\omega _{c}}^{\infty }\,d\omega \,
\frac{\sqrt{\sigma (\omega )\sigma (\omega ')}}{\omega ' - \omega - i0^{+}}
\,b^{\dag }(\omega ')\,\right) \,.
\end{eqnarray}
Here $F(z)$ is analytic except on the branch cut along the real axis
$z > \omega _{c}$ and is given by
\begin{equation}
F(z) = \frac{1}{-\,z + E_{1} - \int_{\omega _{c}}^{\infty }\,d\omega \,
\frac{\sigma (\omega )}{\omega - z}} \,.
\end{equation}

One can explicitly check that the canonical commutation relation,
\begin{equation}
[\,B(\omega ) \,, B^{\dag }(\omega ')\,] = \delta (\omega - \omega ')\,, 
\end{equation}
and the important inversion relation and the Hamiltonian equivalence,
\begin{eqnarray}
&&
b^{\dag }(\omega ) = B^{\dag }(\omega ) + 
\int_{\omega _{c}}^{\infty }\,d\omega' \,
\frac{\sqrt{\sigma (\omega )\sigma (\omega ')}\,F^{*}(\omega ' + i0^{+})}
{\omega - \omega ' + i0^{+}}\,B^{\dag }(\omega ') \,, 
\\ &&
c^{\dag } = -\,\int_{\omega _{c}}^{\infty }\,d\omega \,
\sqrt{\sigma (\omega )}\,F^{*}(\omega + i0^{+})\,B^{\dag }(\omega ) 
\,, 
\\ &&
H = \int_{\omega _{c}}^{\infty }\,d\omega \,\omega \,B^{\dag }(\omega )
B(\omega ) \,.
\end{eqnarray}

The basic reason of integrability is saturation of the unitarity 
relation by "elastic" one;
\begin{equation}
F(\omega + i0^{+}) - F(\omega - i0^{+}) =
2\pi i\,\sigma (\omega )|F(\omega + i0^{+})|^{2} \equiv 
2\pi i\,H(\omega ) \,.
\end{equation}
The quantity $H(\omega )$ is characterized as the overlap between
the prepared state $c^{\dag }|0\rangle $ and the eigenstate
$|\omega\rangle  _{S}$;
\begin{equation}
H(\omega ) =  |\langle 0|c|\omega   \rangle_{S}|^{2} \,.
\end{equation}
In the weak coupling of $\sigma (\omega ) \ll M$, the spectral function
$H(\omega )$ has a Breit-Wigner form as seen from the formula,
\begin{equation}
H(\omega ) = 
\frac{\sigma (\omega )}{(\omega - E_{1} + \Pi (\omega ))^{2}
+ (\,\pi \sigma (\omega)\, )^{2}} 
\,, 
\end{equation}
where $\Pi (\omega )$ is real and given by the dispersion integral,
\begin{equation}
\Pi (z) = {\cal P}\,\int_{\omega _{c}}^{\infty }\,d\omega \,
\frac{\sigma (\omega )}{\omega - z} \,.
\end{equation}

As is well known, there is a simple pole of $F(z)$ in the second Riemann sheet
near the real axis which describes the time evolution in
the form of the exponential decay;
\( \:
e^{-iE_{1}t - \Gamma t/2}
\: \). 
The imaginary part of this pole coincides, in the weak coupling limit, with
the decay rate given by perturbation theory;
\( \:
\Gamma = 2\pi \sigma (E_{1}) 
\: \).

Since the operator solution is known, one can explicitly write down 
many quantities of interest;
for instance the non-decay amplitude of a pure unstable state 
$c^{\dag }| 0 \rangle $ is given by
\begin{equation}
\langle 1|e^{-iHt}|1 \rangle = \langle 0|c\,e^{-iHt}\,c^{\dag }|0 \rangle
= \int_{\omega _{c}}^{\infty }\,d\omega\,\sigma (\omega )\,
|F(\omega + i0^{+})|^{2}\,e^{-i\omega t} 
\,.
\end{equation}
We are primarily interested in the occupation number and 
its time evolution in the cosmological thermal medium.
Let us first recall this quantity in the pure state
\( \:
|\psi (t) \rangle = e^{-iHt}\,|1\rangle 
\: \)
at time $t$,
\begin{equation}
\langle \psi (t)|\,c^{\dag }c\,|\psi (t) \rangle =
\langle 1|\,e^{iHt}\,c^{\dag }c\,e^{-iHt}\,|1 \rangle
=
\langle 1|\,c^{\dag }(t)c(t)\,|1 \rangle \,.
\end{equation}
Here $c(t) = e^{iHt}\,c\,e^{-iHt}$ is the Heisenberg operator and
in this model
\begin{eqnarray}
c^{\dag }(t) = 
-\,\int_{\omega _{c}}^{\infty }\,d\omega \,
\sqrt{\sigma (\omega )}\,F^{*}(\omega + i0^{+})\,e^{i\omega t}\,
B^{\dag }(\omega ) \,.
\end{eqnarray}
More conveniently, it is written in terms of the parent and daughter operators,
\begin{eqnarray}
c^{\dag }(t) &=& 
g(t)\,c^{\dag } + i\,
\int_{\omega _{c}}^{\infty }\,d\omega \,
\sqrt{\sigma (\omega )}\,h(\omega \,, t)e^{i\omega t}\,b^{\dag }(\omega )
\,, 
\\ 
g(t) &=& \int_{\omega _{c}}^{\infty }\,d\omega \,H(\omega )e^{i\omega t}
= \int_{\omega _{c}}^{\infty }\,d\omega \,|\langle 1|\omega  \rangle_{S}|^{2}
\,e^{i\omega t} \,, 
\\ 
h(\omega \,, t) &=& i\,F^{*}(\omega + i0^{+}) - ik(\omega \,, t) \,, 
\hspace{0.5cm} 
\dot{k}(\omega \,, t) = i\,e^{-i\omega t}\,g(t) \,,
\end{eqnarray}
where a condition $k(\omega \,, \infty ) = 0$ is imposed such that
the asymptotic value is
\begin{equation}
h(\omega \,, \infty ) = iF^{*}(\omega + i0^{+}) \,.
\end{equation}
The occupation number in the pure state is thus given by
\begin{equation}
\langle1| c^{\dag }(t)c(t)|1 \rangle = |g(t)|^{2} \,.
\end{equation}

There is a very useful way \cite{jmy-96} to compute the basic function $g(t)$.
One can use the analytic property of $F(z)$ to express this function
as a sum of two contour integrals as shown in Fig.1; 
the first one encircling the pole in the second sheet
($C_{0}$) and the second continuous integral along a complex path $C_{1}$;
\begin{eqnarray}
g(t) = \frac{1}{2\pi i}\,\left( \,  \int_{C_{0}} + \int_{C_{1}} \,\,\right)
\,dz\,F(z)e^{iz t} \equiv g_{0}(t) + g_{1}(t)
\,.
\end{eqnarray}
Physically, this second continuous contribution $g_{1}(t)$
gives the off-shell effect, while the pole contribution $g_{0}(t)$
essentially gives the on-shell effect.
Both terms decrease as $t\rightarrow \infty $, but
the $C_{1}$ integral has a power dependence \cite{jmy-97-1} 
in contrast to the exponential form of the pole term; 
\( \:
g_{1}(t) \propto t^{-\alpha - 1} \,, 
\: \)
where $\alpha $ is related to the threshold behavior of the spectral
function,
\( \:
\sigma (\omega ) \approx c(\omega - \omega _{c})^{\alpha } \,,
\: \)
near $\omega = \omega _{c}$.

There is an equivalent and more intuitive way to separate the on-shell
and the off-shell contributions. For this we go back to the $\omega $
integral along the real axis,
\begin{equation}
g(t) = \int_{\omega _{c}}^{\infty }\,d\omega \,
\frac{\sigma (\omega )}{(\omega - E_{1} + \Pi (\omega ))^{2}
+ (\,\pi \sigma (\omega)\, )^{2}}\, e^{i\omega t} \,.
\end{equation}
Assuming the weak coupling, 
\( \:
\Pi (\omega ) \ll E_{1} \,,
\: \)
one separates the region of integration into the two parts,
one around the pole,
\( \:
\omega \approx E_{1} + \Pi (E_{1}) \,, 
\: \)
and the rest of the region which is dominated near the
threshold, $\omega = \omega _{c}$, for a large time.
This approximation gives
\begin{equation}
g(t) \approx \exp [\,iE_{1}t - \pi \sigma (E_{1})t\,]
+ \frac{1}{E_{1}^{2}}\,\int_{\omega _{c}}^{E_{c}}\,d\omega \,
\sigma (\omega )\,e^{i\omega t} \,,
\end{equation}
where $E_{c}$ is a physical cutoff scale of order $E_{1}$.
The first term is the on-shell contribution, while the second is
the off-shell contribution.
In computation of physical quantities in thermal medium
there are other energy scales such as the temperature $T$, which may replace
this cutoff by the factor $e^{- \omega /T}$.
The important message is that the off-shell effect at late times
is determined by the $\omega $ integration near the threshold region.

Extention to the many channel problem is straightforward. 
We denote the channel by an index $i$ and write the decay interaction as
\begin{equation}
\int_{\omega _{c}}^{\infty }\,d\omega \,\sum_{i}\,\sqrt{\sigma _{i}(\omega )}
\,\left( b_{i}^{\dag }(\omega )c + c^{\dag }b_{i}(\omega ) \right) \,.
\end{equation}
Only the key formulas are quoted;
\begin{eqnarray}
&& \hspace*{2cm}
B_{i}^{\dag }(\omega ) = b_{i}^{\dag }(\omega ) 
\nonumber 
\\ && 
+ F(\omega + i0^{+})\,
\left( \,-\,\sqrt{\sigma _{i}(\omega )}\,c^{\dag } +
\int_{\omega _{c}}^{\infty }\,d\omega '\,\sum_{k}\,
\frac{\sqrt{\sigma _{i}(\omega )\,
\sigma_{k} (\omega ')}}{\omega ' - \omega - i0^{+}}\,b_{k}^{\dag }(\omega' )
\,\right)\,, 
\\ && \hspace*{1cm} 
F(z) = \frac{1}{-\,z + E_{1} - \int_{\omega _{c}}^{\infty }\,d\omega \,
\frac{\sum_{i}\sigma _{i}(\omega )\,}{\omega - z}\,} \,, 
\\ &&
c^{\dag }(t) = 
g(t)\,c^{\dag } + i\,\int_{\omega _{c}}^{\infty }\,d\omega \,h(\omega \,, t)
\,e^{i\omega t}\,\sum_{i}\,\sqrt{\sigma_{i} (\omega )}\,b_{i}^{\dag }(\omega )
\,, 
\\ && 
b_{i}^{\dag }(\omega \,, t) = e^{iHt}\,b_{i}^{\dag }\,e^{-iHt}
= e^{i\omega t}\,\left( \,\sqrt{\sigma _{i}(\omega )}\,
ih(\omega \,, t)\,c^{\dag } + b_{i}^{\dag }(\omega )\,\right) \nonumber 
\\ && 
+ \, \sqrt{\sigma _{i}(\omega )}
\int_{\omega _{c}}^{\infty }\,d\omega '\,
\frac{\sum_{k}\sqrt{\sigma _{k}(\omega ')}b_{k}^{\dag }(\omega ')}
{\omega - \omega ' + i0^{+}}\,
\left( \,ih(\omega \,, t)e^{i\omega t} - ih(\omega' \,, t)e^{i\omega' t}
\,\right)
\,,
\\ && \hspace*{1cm} 
g(t) = \int_{\omega _{c}}^{\infty }\,d\omega \,\sum_{i}\,\sigma _{i}(\omega )
\,|F(\omega + i0^{+})|^{2}\,e^{i\omega t} \,, 
\\ && \hspace*{1cm} 
h(\omega \,, t) = i\,\left( \,F^{*}(\omega + i0^{+}) 
- k(\omega \,, t)\,\right) \,, 
\\ && \hspace*{1cm} 
k(\omega \,, t) = \frac{1}{2\pi i}\,
\int_{C_{0} + C_{1}}\,dz\,\frac{F(z)}{z - \omega }\,e^{i(z - \omega )t}
\,. 
\end{eqnarray}
Note that both the analytic function $F(z)$ and the basic $g(t)$
are determined by the total strength of the spectral function,
\begin{equation}
\sigma (\omega ) = \sum_{i}\,\sigma _{i}(\omega ) \,.
\end{equation}

We now consider the field theory model of unstable particles, denoting
the parent particle by $c$ and two-body daughter particles by 
$b_{1}\,b_{2}$.
The decay interaction of the unstable particle of a momentum $\vec{q}$
is given by the Hamiltonian;
\begin{eqnarray}
&&
\hspace*{-1cm}
H_{{\rm int}} = 
\int\,\frac{d^{3}k_{1}d^{3}k_{2}}{(2\pi )^{6}}\,
(2\pi )^{3}\delta ^{3}(\vec{q} - \vec{k}_{1} - \vec{k}_{2})\,
\frac{g}{\sqrt{8\omega _{1}\omega _{2}\omega _{q}}}\,
\left(\, b_{1}^{\dag }(\vec{k}_{1})b_{2}^{\dag }(\vec{k}_{2})c(\vec{q})
+ ({\rm h.c.}) \,\right)
\,, \nonumber \\ &&
\end{eqnarray}
with $g$ some coupling constant.
We may then identify the decay product operator and the spectral function
as
\begin{eqnarray}
&&
\hspace*{1cm} 
\sum_{i}\,\sqrt{\sigma_{i} (\omega \,, \vec{q})}
\,b_{i}^{\dag }(\omega \,, \vec{q})
= \nonumber 
\\ && \hspace*{-1cm} 
\int\,\frac{d^{3}k_{1}d^{3}k_{2}}{(2\pi )^{6}}\,
(2\pi )^{3}\delta ^{3}(\vec{q} - \vec{k}_{1} - \vec{k}_{2})\,
\delta (\omega - \omega _{1} - \omega _{2})\,
\frac{g}{\sqrt{8\omega _{1}\omega _{2}\omega _{q}}}\,
b_{1}^{\dag }(\vec{k}_{1})b_{2}^{\dag }(\vec{k}_{2})
\,, \label{composite operator} 
\\ && 
\hspace*{1cm} 
\sigma  (\omega \,, \vec{q}) = 2\pi \,\sum_{i}\,\sigma _{i}(\omega \,, \vec{q})
= 
\nonumber 
\\ &&
\int\,\frac{d^{3}k_{1}d^{3}k_{2}}{(2\pi )^{3}}\,
\delta ^{3}(\vec{q} - \vec{k}_{1} - \vec{k}_{2})\,
\delta (\omega - \omega _{1} - \omega _{2})\,
\frac{g^{2}\,\prod_{i}\,(2\omega _{i})^{F_{i}}}{8\omega _{1}\omega _{2}
\omega _{q}}\,\sum_{{\rm spins}}\,|{\cal M}|^{2} \,, 
\end{eqnarray}
where $F_{i} = 1$ for fermions and $F_{i} = 0$ for bosons.
The decay amplitude ${\cal M}$ should be given separately in specific
decay models.
This can be thought of an extention of the discretely many (and finite) 
decay channel problem to the continuously many (and infinite) channel
problem.

The field theory model of unstable particle decay 
is not exactly solvable, because
the commutator among the decay product operator,
\begin{eqnarray}
&&
[\, b_{1}b_{2} \,, b_{1}^{\dag }b_{2}^{\dag } \,] = 
1 \pm  (\,b^{\dag }_{1}b_{1} + b^{\dag }_{2}b_{2} \,)
\end{eqnarray}
($\pm $ referring to the boson or fermion pair),
is not the cannonical one 
\( \:
[\,b \,, b^{\dag }\,] = 1
\: \)
as in the quantum mechanical model.
Thus, the composite operators $b_{i}^{\dag }(\omega \,, \vec{q})$
introduced by eq.(\ref{composite operator}) do not obey the canonical
commutation relation so crucial to the integrability.
But the important case in which the bilinear term in the right hand side
of the commutator can be neglected and
the replacement is made,
\begin{equation}
[\, b_{1}b_{2} \,, b_{1}^{\dag }b_{2}^{\dag } \,]  \:\rightarrow  \: 1 \,,
\end{equation}
is identical to the solvable model in quantum mechanics.
This occurs in the circumstance under which the thermal medium is 
very far away from the degeneracy limit. In this case the occupation number 
$f_{i}$ which is the expectation value of the operators
\( \:
b_{i}^{\dag }b_{i} 
\: \)
in the thermal medium is very small ($f_{i} \ll 1$), 
and one may neglect the bilinear term in the commutator above.
Even in the dense, hot early universe of the standard cosmology
the low occupation number is realized.
We shall thus fully exploit this approximation in application to
cosmology.

From the definition of the spectral function a relation to the decay
rate follows;
\begin{equation}
\sigma (\omega _{q} \,, \vec{q}) = \frac{M}{\omega _{q}}\,\frac{\Gamma }
{2\pi } \,,
\end{equation}
on the mass shell with $\omega = \omega _{q} = \sqrt{M^{2} + q^{2}}$.
The factor $\frac{M}{\omega _{q}}$ represents the time dilatation
effect.
Off the mass shell,
\begin{equation}
\sigma (\omega \,, \vec{q}) = \frac{M}{\omega _{q}}\,
\sigma (\sqrt{\omega ^{2} - \vec{q}^{2}})
\,, 
\end{equation}
where the spectral function in the right hand side $\sigma (\omega )$
is the one in the rest frame.

A choice of the decay model corresponds to a particular form of the
spectral function.
For instance, 
in the fermion-pair (of equal mass $m$) decay of a scalar boson 
$\varphi \rightarrow \psi \bar{\psi }$ described by
a Lagrangian density of
\( \:
{\cal L} = g\,\varphi \bar{\psi }\psi \,, 
\: \)
\begin{equation}
\sigma (\omega \,, \vec{k}) = 
\frac{g^{2}}{16\pi ^{2}\,\sqrt{k^{2} + M^{2}}}\,(\omega ^{2} - k^{2})\,
\left(1 - \frac{4m^{2}}{\omega ^{2} - k^{2}}\right)^{3/2} \,.
\label{spectral for f-pair decay} 
\end{equation}
A more general, convenient parametrization of the spectral function 
that becomes adequate 
in the temperature range of $T \gg 2m$ (the threshold for the decay
product pair) is given, using the decay rate $\Gamma $;
\begin{equation}
\sigma (\omega \,, \vec{k}) 
= \frac{\Gamma }{2\pi }\,\frac{M}{\sqrt{k^{2} + M^{2}}}
\,\frac{(\omega ^{2} - k^{2})^{\alpha /2}}{M^{\alpha }} \,.
\label{spectral form} 
\end{equation}
For instance, the gauge X boson decay into a fermion-pair as well as
the scalar $X$ boson decay given by
(\ref{spectral for f-pair decay}) has this form with
$\alpha =2$, while the Higgs X boson decay into a boson-pair has
this form with $\alpha = 0$.

In reality, the spectral function in any non-trivial field theory model
is complicated beyond the lowest order of perturbation.
But it turns out that what is important for our subsequent analysis 
is the on-shell value of the spectral function given by the decay rate
$\Gamma $ and its behavior near the decay threshold, hence the
parameter $\alpha $ in addition to the decay rate.
The value of $\alpha $ is dictated by the unitarity relation for the opening
channel, thus is essentially of kinematical origin.
Both intermediate and late time behaviors of the decay, and important 
temperature dependent off-shell
effects are described by these two parameters.
Hence in a sense the detailed specification of a field theory for the decay
is unnecessary.

The occupation number of the unstable particle decaying in vacuum is
thus given by $|g(\vec{q} \,, t)|^{2}$ with
\begin{equation}
g(\vec{q} \,, t) = \frac{M}{\sqrt{M^{2} + q^{2}}}\,
\int_{m_{1} + m_{2}}^{\infty }\,d\omega \,\sigma (\sqrt{\omega ^{2} - q^{2}})
\,|F(\omega + i0^{+} \,, \vec{q})|^{2}\,e^{i\omega t} \,,
\end{equation}
where $m_{i}$ are masses of decay product particles.

We should mention a limitation of our approach.
We neglected in our model Hamiltonian
the interaction of unstable particles with the medium, 
except the decay interaction.
This guarantees a coherence of the decay and its inverse
interaction and makes it
easy to treat the environment effect on the decay process.
There may however be an important class of thermal interactions on the
unstable particle; the scattering off thermal light particles.
If the thermal interaction of this sort is included, it gives rise
to an additional term to the spectral function.
In terms of the coupling strength this is a higher order effect
of order $\alpha ^{2}\,T $ compared to the decay rate of order
$\alpha M$. However, this might contribute in the off-shell region.
We shall go back to this effect when we discuss the off-shell effect
in thermal medium in Section 4.

In the field theory of unstable particle decay one needs to perform 
renormalization.
The method of renormalization is explained elsewhere \cite{jmy-97-1} 
and here we shall write for simplicity
quantities without renormalization, since in lowest order of perturbation
renormalization is straightforward.

\vspace{0.5cm} 
\section{Generalized Boltzmann equation}
\vspace{0.5cm} 
\hspace*{0.5cm} 
In order to discuss the decay process of unstable particles that
occur in the cosmic thermal medium, one must take into account
the presence of medium and incorporate the inverse process that
creates the unstable particle.
A thermal environment is described by the density matrix denoted here
by $\rho _{i}$
and one has to consider how the decay proceeds in this mixed state.
For the time being we assume
that the change of the thermal environment
is minor and the back reaction of the environment change against the decay
and its inverse process is negligible.
Later when we extend our analysis to baryogenesis, we incorporate
a relevant effect of the environment change.

In the mixed state the occupation number of an excited level is
\begin{equation}
f(t) = {\rm tr}\,\left( c^{\dag }(t)c(t)\,\rho _{i} \right)
\,.
\end{equation}
It is not difficult to show from the operator solution
that this quantity obeys the first order differential equation;
\begin{eqnarray}
&& \hspace*{1cm}
\frac{df}{dt} - 2\Re\,\frac{\dot{g}}{g}\,f =
\nonumber \\ &&
i\,\int_{\omega _{c}}^{\infty }\,d\omega \,
\sqrt{\sigma (\omega )}\,
g^{*}\,\left( \,
g + i\omega he^{i\omega t} - \frac{\dot{g}}{g}he^{i\omega t}\,\right)\,
\langle b^{\dag }(\omega )c \rangle_{i}
+ ({\rm c.c})
\nonumber 
\\ && 
+\,
\int_{\omega _{c}}^{\infty }\,d\omega \,
\int_{\omega _{c}}^{\infty }\,d\omega' \,
\sqrt{\sigma (\omega )\sigma (\omega ')}\;
\left( \,gh^{*}(\omega '\,, t)e^{-i\omega t} + g^{*}h(\omega \,, t)
e^{i\omega ' t} 
\right. 
\nonumber 
\\ && \hspace*{-1cm}
\left.\,
+\, i(\omega - \omega ')h(\omega \,, t)h^{*}(\omega '\,, t)
- 2\Re \frac{\dot{g}}{g}\,h(\omega \,, t)h^{*}(\omega ' \,, t)\,\right)
e^{i(\omega - \omega ')t}
\,\langle b^{\dag }(\omega )b(\omega ') \rangle_{i} \,.
\end{eqnarray}
We used the notation,
\begin{equation}
\langle A \rangle_{i} \equiv {\rm tr}\,(A\rho _{i}) \,.
\end{equation}
Although one can write $f(t)$ in an integrated form, this differential
equation is more useful when one incorporates effect of the cosmic expansion.
Another advantage of this form is that the initial state dependence
via
\( \:
\langle c^{\dag }c \rangle_{i}
\: \)
is eliminated in favor of the occupation number $f(t)$ at any time $t$.
It is however convenient not to eliminate the initial state dependence
of environment variables,
\( \:
\langle b^{\dag }(\omega )b(\omega ') \rangle_{i} \,, 
\: \)
when we later incorporate the environment change.

We take the uncorrelated initial state satisfying
\begin{equation}
\langle c^{\dag }c \rangle_{i} = f(0) \,, \hspace{0.5cm} 
\langle c^{\dag }b(\omega ) \rangle_{i} = 0 \,,
\hspace{0.5cm} 
\langle b^{\dag }(\omega )b(\omega ') \rangle_{i}
= f_{i}(\omega )\,\delta (\omega - \omega ') \,.
\end{equation}
Although the unstable particle may or may not be in thermal equilibrium
at much higher temperatures in the earlier epoch, 
the choice of the uncorrelated initial state seems
reasonable, because there exists a time lag between the unstable and
the decay product particles due to different interaction among themselves
and with the rest
of the environment (not written in the Hamiltonian above).
We imagine that when unstable particles are about to decay, the decay
interaction cannot keep pace with fast thermal interaction in the bulk
of medium, hence we assume that the unstable particle has 
no thermal contact with the medium when
we start calculation of the abundance evolution.
In the end of the next section we however estimate how the scattering
off thermal particles may affect the decay law in thermal medium.

Introducing the rate defined by
\begin{equation}
\Gamma (t) \equiv 
-\,2 \,\Re \frac{\dot{g}}{g} = -\,\frac{d}{dt}\ln |g(t)|^{2}
\end{equation}
(which reduces to a constant rate $\Gamma $ in the pole dominance
approximation), one has
\begin{eqnarray}
&&
\hspace*{2cm}
\frac{df}{dt} + \Gamma (t)\,f =
\nonumber 
\\ &&
\int_{\omega _{c}}^{\infty }\,d\omega \,\sigma (\omega )\,
\left( \,2\Re (g(t)h^{*}(\omega \,, t)e^{-i\omega t}) +
\Gamma (t)\,|h(\omega \,, t)|^{2}\,\right)\,
f_{i}(\omega ) 
\,. \label{occupation-n evol} 
\end{eqnarray}

It is instructive to first give the time evolution in the narrow width
approximation and then explain an improved approximation incorporating
the $C_{1}$ integral.
Using
\begin{equation}
g_{0}(t) \approx e^{-\Gamma t/2 + iE_{1}t} \,, 
\end{equation}
one has
\begin{eqnarray}
&&
\hspace*{2cm}
2\Re (g_{0}h_{0}^{*}e^{-i\omega t}) 
- 2\Re \frac{\dot{g_{0}}}{g_{0}}\,|h_{0}|^{2}
\approx  
\frac{1}{(\omega - E_{1})^{2} + \frac{\Gamma^{2}}{4}}
\nonumber 
\\ &&
\cdot \left( \,\Gamma(\,1 - e^{-\Gamma t/2}\,
\cos (\omega - E_{1})t\,) +
2e^{-\Gamma t/2}(\omega - E_{1}) \sin (\omega - E_{1})t \,\right) \,.
\end{eqnarray}
There are oscillatory terms of frequency of $\approx 1/(\omega - E_{1})$,
but they are averaged out by the $\omega $ integration.
From eq.(\ref{occupation-n evol}) 
the fundamental equation in thermal medium is found to be
\begin{equation}
\frac{df}{dt}  = -\,\Gamma\,
\left( f - f_{i}(E_{1}) \right)
\,,
\end{equation}
in the narrow width approximation.
Thus, this equation describes the relaxation towards the environment
value by the constant rate $\Gamma $.

The evolution equation (\ref{occupation-n evol})
correctly describes the relaxation process of unstable particle in medium
beyond the narrow width approximation.
Since $g(t) \rightarrow 0$ and $h(\omega \,, t)$
approaches an equilibrium value as $t\rightarrow \infty $, 
the stationary value of the occupation number is
\begin{equation}
f(t) \:\rightarrow  \: \int_{\omega _{c}}^{\infty }\,d\omega \,\sigma (\omega )
\,|F(\omega + i0^{+})|^{2}\,f_{i}(\omega ) 
\equiv f_{\infty}
\,.
\end{equation}
If one takes a thermal distribution for the initial density matrix,
\begin{equation}
f_{\infty } = \int_{\omega _{c}}^{\infty }\,d\omega \,
\frac{\sigma (\omega )}{(\omega - E_{1} + \Pi (\omega ))^{2}
+ (\pi \sigma (\omega))^{2}} 
\,\frac{1}{e^{\beta \omega } - 1} \,.
\end{equation}
The narrow width approximation $\sigma(\omega ) \ll E_{1}$ of this gives
the well-known result;
\begin{equation}
f_{\infty } \approx \frac{1}{e^{\beta E_{1}} - 1} \,,
\end{equation}
but this form is only approximate and is not a good one in low temperatures.
In low temperatures of
\( \:
T \ll E_{1} \,, 
\: \)
the $\omega $ integral for $f_{\infty }$
is dominated by the contribution near the threshold; 
taking the form of the threshold rise
\( \:
\sigma (\omega ) \approx c\,(\omega - \omega _{c})^{\alpha }
\: \)
gives 
\begin{equation}
f_{\infty } \approx \frac{c}{(E_{1} - \omega _{c})^{2}}\,
\int_{\omega _{c}}^{\infty }\,d\omega \,\frac{(\omega - \omega _{c})^{\alpha }}
{e^{\beta \omega } - 1} \,.
\end{equation}
When $T \ll \omega _{c}$ or $T \gg \omega _{c}$, this further simplifies to
\begin{equation}
f_{\infty } \approx \frac{c\,\Gamma (\alpha + 1)}{(E_{1} - \omega _{c})^{2}}\,
e^{-\beta \omega _{c}}\,T^{\alpha + 1} \,,
\hspace{0.5cm} {\rm for} \; T \ll \omega _{c} \,, 
\end{equation}
where $\Gamma (x)$ is the Euler's gamma function, and
\begin{equation}
f_{\infty } \approx \frac{c\,\zeta (\alpha + 1)
\Gamma (\alpha + 1)}{(E_{1} - \omega _{c})^{2}}\,T^{\alpha + 1} \,,
\hspace{0.5cm} {\rm for} \; E_{1} \gg T \gg \omega _{c} \,,
\end{equation}
with $\zeta (x)$ the Riemann's zeta function.
The Boltzmann suppressed temperature dependence of the occupation number
\( \:
e^{-E_{1}/T}
\: \)
near $T = E_{1}$ is thus changed to the power behaved
\( \:
T^{\alpha + 1} 
\: \)
at low temperatures \cite{jmy-96}.
What caused this big change is that the full Breit-Wigner shape is cut
off effectively at the temperature $T$, since $T \ll $
the center location of the Breit-Wigner function.

We now turn to the unstable particle decay.
The occupation number for a mode $\vec{k}$ is
\begin{eqnarray}
&& 
f(\vec{k} \,, t) = |g(\vec{k} \,, t)|^{2}\,f_{{\rm th}}(\omega _{k})
+ \int\,d\omega \,|h(\omega \,, \vec{k} \,, t)|^{2}
\nonumber \\ && \hspace*{-1.5cm} 
\int\,\frac{d^{3}k_{1}d^{3}k_{2}}{(2\pi )^{3}}\,
\frac{g^{2}\,\prod_{i}\,(2\omega _{i})^{F_{i}}}{8k_{1}k_{2}\omega _{k}}\,
\sum_{{\rm spins}}\,|{\cal M}|^{2}
\,f_{{\rm th}}(k_{1})f_{{\rm th}}(k_{2})
\,\delta ^{3}(\vec{k} - \vec{k}_{1} - \vec{k}_{2})\,\delta 
(\omega - k_{1} - k_{2}) 
\,. \nonumber \\ &&
\end{eqnarray}
We took the massless particle for decay products such that
\( \:
\omega _{i} = |\vec{k}_{i}| \,.
\: \)
The stationary limit of the occupation number is then
\begin{eqnarray}
&& \hspace*{0.5cm} 
f_{\infty }(\vec{k}) = \int\,d\omega \,
|h(\omega \,, \vec{k} \,, \infty )|^{2}\,
\int\,\frac{d^{3}k_{1}d^{3}k_{2}}{(2\pi )^{3}}\,
\nonumber \\ && \hspace*{-1cm}
\frac{g^{2}\,\prod_{}\,_{i}\,(2\omega _{i})^{F_{i}}}{8k_{1}k_{2}
\omega _{k}}\,\sum_{{\rm spins}}\,|{\cal M}|^{2}\,
f_{{\rm th}}(k_{1})f_{{\rm th}}(k_{2})
\,\delta ^{3}(\vec{k} - \vec{k}_{1} - \vec{k}_{2})\,\delta 
(\omega - k_{1} - k_{2}) 
\,.
\end{eqnarray}

In the low occupation number limit this is further simplified since
\begin{eqnarray*}
f_{{\rm th}}(k_{1})f_{{\rm th}}(k_{2}) = e^{-\beta (k_{1} + k_{2})}
\,, 
\end{eqnarray*}
which is replaced by $e^{-\beta \omega }$ in the above integrand.
It is thus found that
\begin{eqnarray}
&&
f_{\infty }(\vec{k}) \approx \int_{k}^{\infty }\,d\omega \,
\frac{\sigma (\omega \,, \vec{k})\,e^{-\beta \omega }}
{(\omega - \omega _{k})^{2} + \pi ^{2}\sigma^{2}(\omega \,, \vec{k})}
\,.
\end{eqnarray}
Separating the pole and the threshold regions for this $\omega $ integral,
one has
\begin{eqnarray}
&&
f_{\infty }(\vec{k}) \approx e^{-\beta \omega _{k}} +
\frac{\Gamma \,k^{\alpha + 1}}{2\pi \,M^{\alpha + 2}}\,
\int_{1}^{\infty }\,dx\,(x^{2} - 1)^{\alpha /2}\,e^{-\beta k\,x}
\,.
\label{late-time distribution 1} 
\end{eqnarray}
The last integral in this equation is given by the modified Bessel function,
and in the large $k/T$ limit it behaves as
\begin{equation}
k^{\alpha /2}\,e^{- k/T} \,.
\label{late-time distribution} 
\end{equation}
This has a narrower spread of the momentum than of order $\sqrt{MT}$ 
for the Maxwell-Boltzmann
distribution, and its average is of order $T$
(actually $(\frac{\alpha }{2} + 3)\times T$ using the distribution,
(\ref{late-time distribution})).
We shall later use this fact to simplify the momentum dependence of
the effective decay rate at late times.
We plot the momentum distribution given by eq.(\ref{late-time distribution 1})
in Fig.2 to compare with the on-shell distribution relevant at
low temperatures; $e^{-\,\sqrt{k^{2} + M^{2}}/T}$.

The total number density $n(t)$ is then obtained by summing the occupation
number over modes of various particle momentum
$\vec{k}$. In an isotropic medium it is
\begin{equation}
n(t) = \frac{1}{2\pi ^{2}}\,\int_{0}^{\infty }\,dk\,k^{2}f(k\,, t) \,.
\end{equation}
In general, various quantities in eq.(\ref{occupation-n evol}) depends
on $\vec{k}$. This dependence is traced to the boost of the parent unstable
particle.
We shall often suppress the $\vec{k}$ dependence, unless otherwise
there is a confusion.
The stationary number density $n_{\infty }$ 
is determined using $f_{\infty }$ for $f(\vec{k} \,, t)$;
in the two-particle decay of equal mass $m$
\begin{eqnarray}
&& \hspace*{-1.5cm}
n_{\infty } = \frac{1}{2\pi ^{2}}\,\int_{0}^{\infty }\,dk\,k^{2}\,
\int_{\sqrt{k^{2} + 4m^{2}}}^{\infty }\,
d\omega \,\frac{\sigma (\omega \,, k)}{(\omega - 
\sqrt{k^{2} + M^{2}})^{2} + \pi ^{2}\,\sigma^{2}(\omega \,, k)}
\,\frac{1}{e^{\omega /T} - 1} \,.
\label{stationary n-density} 
\end{eqnarray}
Throughout this work we ignored a small part of $\Pi (\omega )$;
the residual $\Pi (\omega )$ that remains after renormalization.

In the low temperature range of
\( \:
M \gg T \gg 2m \,
\: \)
the time dilatation effect is negligible and
\begin{equation}
n_{\infty } \approx \frac{\Gamma }{4\pi ^{3}}\,\int_{0}^{\infty }\,d\omega \,
\frac{1}{e^{\beta \omega } - 1}\,\int_{0}^{\omega }\,dk\,k^{2}\,
\frac{(\omega ^{2} - k^{2})^{\alpha /2}}{M^{\alpha + 2}} \,.
\end{equation}
It is then analytically calculated as
\begin{eqnarray}
&&
\frac{n_{\infty } }{T^{3}} 
\approx A(\alpha )\,\frac{\Gamma }{M}\,(\frac{T}{M})^{\alpha  + 1} \,,
\hspace{0.5cm} 
A(\alpha ) = 
\frac{\zeta (\alpha + 4)\Gamma (\alpha + 4)\Gamma (\frac{\alpha }{2} + 1)}
{16\pi ^{2}\,\sqrt{\pi }\,\Gamma (\frac{\alpha }{2} + \frac{5}{2})} \,.
\label{off-shell asymptotic} 
\end{eqnarray}

In Fig.3 we plotted the stationary number density, 
eq.(\ref{stationary n-density}) combined with the spectral function
(\ref{spectral form}), as a function of
$T/M$ for two values of the decay rate $\Gamma $.
Our numerical computation supports that to a good accuracy the stationary
number density $n_{\infty }$ is given by a sum of the pole and the threshold
contribution;
\begin{equation}
n_{\infty } = \frac{1}{2\pi ^{2}}\,\int_{0}^{\infty }\,dk\,
\frac{k^{2}}{e^{\sqrt{k^{2} + M^{2}}/T} - 1} +
A(\alpha )\,\frac{\Gamma }{M}(\frac{T}{M})^{\alpha + 1}\,T^{3} \,.
\label{approximate stationary f} 
\end{equation}
This formula is accurate 
for any temperature $T $ less than $ M$ if the decay
rate $\Gamma /M$ is small enough. But at higher temperatures
the first on-shell term in eq.(\ref{approximate stationary f}) alone
is accurate and
the off-shell power term $\propto T^{\alpha + 4}$ in this equation
should be discarded.

One may define the equal temperature $T_{eq}$ as the one at which
the two contributions in this equation, the pole and the threshold
contributions, become equal. How this equal temperature depends on
the rate $\Gamma /M$ is important, and it is shown in Fig.4 for
two values of $\alpha $.
With the on-shell contribution given by
\begin{equation}
n_{\infty } \approx (\frac{MT}{2\pi})^{3/2}\,e^{-M/T} \,, 
\end{equation}
the equation that determines $T_{{\rm eq}}$ is
\begin{equation}
x^{\alpha + \frac{5}{2}}\,e^{-x} = (2\pi )^{3/2}\,A(\alpha )\frac{\Gamma }{M}
\,, \hspace{0.5cm} x \equiv \frac{M}{T_{{\rm eq}}} \,.
\end{equation}
for a small $\Gamma /M$.
A rough analytic estimate in the $\Gamma \rightarrow 0$ limit would be
\begin{eqnarray}
&&
\frac{T_{{\rm eq}}}{M} \approx 
\left( \,\ln \frac{M}{(2\pi )^{3/2}A(\alpha )\Gamma }
+ (\alpha + \frac{5}{2})\,\ln \ln \frac{M}{(2\pi )^{3/2}A(\alpha )\Gamma }\,
\right)^{-1}
\,,  \label{transition temperature} 
\end{eqnarray}
but this expression is accurate only for a very small $\frac{\Gamma }{M}$,
for instance,
\( \:
\frac{\Gamma }{M} < 10^{-4} 
\: \)
for $\alpha = 0$.

We now turn to the time evolution of the occupation number and the
number density.
The approach towards the stationary value is governed by the large time
limit of $g(t)$. In the pole dominance approximation the relaxation
is exponential in time and fast. 
On the other hand, the true late time behavior
is the power law, and this makes 
complete relaxation slower.\cite{jmy-96}

The effect of the cosmological expansion is readily incorporated
for the number density $n$
by adding the $3H\,n$ term where the Hubble rate
\begin{equation}
H = \frac{\dot{a}}{a} = \frac{1}{2t} 
\,.
\end{equation}
We first give the result in the narrow width approximation;
\begin{eqnarray}
&&
\frac{dn}{dt} + 3\frac{\dot{a}}{a}\,n = - \,\overline{\Gamma }\,
(n - n^{{\rm th}}(T)) \,, 
\\ &&
n^{{\rm th}}(T) = \frac{T^{3}}{2\pi ^{2}}\,\int_{0}^{\infty }\,dx\,
\frac{x^{2}}{e^{\sqrt{x^{2} + M^{2}/T^{2}}} - 1} \,.
\end{eqnarray}
We introduced an averaged quantity over the momentum 
in order to simplify the mode integral;
\begin{equation}
\overline{\Gamma } = \frac{1}{2\pi ^{2}\,n(t)}
\,\int_{0}^{\infty }\,dk\,k^{2}\,\Gamma _{k}\,f(k \,, t) \,.
\end{equation}
In low enough temperatures this averaging is indeed simple;
\( \:
\overline{\Gamma } \approx \Gamma \,, 
\: \)
because the time dilatation effect given by
\( \:
M/\sqrt{M^{2} + k^{2}} \approx 1 - \frac{k^{2}}{2M}
\: \)
is negligible for the slow motion of parent particles.
The equation for the number density should be combined with
the well known temperature-time relation in the radiation dominated 
universe,
\begin{equation}
t = \frac{dm_{{\rm pl}}}{T^{2}} \,, \hspace{0.5cm} 
\frac{\dot{a}}{a} = -\,\frac{\dot{T}}{T} \,, \hspace{0.5cm} 
d = \sqrt{\frac{45}{16\pi ^{3}\,N}} \,, 
\label{temperature-time rel} 
\end{equation}
with $N$ the number of massless species contributing to the energy density.

In the low temperature region of $T < M$ the equation is further
simplified by using the Maxwell-Boltzmann distribution function 
of the zero chemical potential,
\begin{equation}
e^{- (M +\frac{k^{2}}{2M})/T} \,.
\end{equation}
A convenient evolution equation is obtained using the dimensionless 
quantities,
\begin{eqnarray}
&&
Y \equiv \frac{n}{T^{3}} \,, \hspace{0.5cm} 
u\equiv \frac{M}{T} = \sqrt{\frac{M^{2}t}{dm_{{\rm pl}}}} \,, 
\hspace{0.5cm} 
\tau \equiv \Gamma t \,.
\end{eqnarray}
The time scale of variation is given by the lifetime $\Gamma^{-1}$,
hence it is useful to use the dimensionless time $\tau $.
The evolution equation is then
\begin{eqnarray}
&&
\frac{dY}{d\tau } = -\,(\,Y - S\,) \,, \hspace{0.5cm} 
S(u) = (\frac{u}{2\pi })^{3/2}\,e^{-u} \,.
\end{eqnarray}
The source term varies as
\( \:
S\left( \sqrt{\frac{\tau }{\eta }}\right)
\,, 
\: \)
with
\begin{equation}
\eta = \frac{dm_{{\rm pl}}\Gamma }{M^{2}}  \,, \hspace{0.5cm} 
u  = \sqrt{\frac{\Gamma t}{\eta }} =
\sqrt{\frac{\tau }{\eta }}\,.
\label{time-temperature relation} 
\end{equation}
Thus, $S$ as a function of $\tau $ has a maximum around $\tau = \eta $.
The meaning of the quantity $\eta $ is the decay rate $\Gamma $
divided by the Hubble rate $H$ at the temperature $T = M$.

One can readily integrate this equation.
The solution to the differential equation for the yield $Y$ is
\begin{equation}
Y(\tau ) = \int_{\tau _{i}}^{\tau }\,d\tau'\,
S\left( \sqrt{\frac{\tau' }{\eta }}\right)\,
e^{-\,(\tau - \tau ')} + Y(\tau _{i})\,e^{-\,(\tau - \tau _{i})} \,.
\end{equation}
The first term gives the yield created from the thermal medium.
For $\tau \gg 1$, namely at times much larger than the lifetime,
$t \gg 1/\Gamma $, the dominant region of the $\tau '$ integral is
$\tau - \tau ' < 1$ and the integral gives
\begin{equation}
Y(\tau ) \approx S\left( \sqrt{\frac{\tau }{\eta }}\right)
\,.
\end{equation}
This is in general valid unless 
\( \:
\eta < \eta _{{\rm cr}} \,, 
\: \)
where the critical value $\eta _{{\rm cr}} = \frac{1}{12}$.
The yield $Y$ thus roughly follows the thermal value $S(T)$.

There is some lesson one can learn on the more general case of
the mode independent $\Gamma (t) = -\,\frac{d}{dt}\ln |g|^{2}$,
from this calculation in the pole dominance approximation. 
It is obvious that even for the more general case  
the general solution to the mode-summed form of eq.(\ref{occupation-n evol})
extended to the expanding universe is given by
\begin{eqnarray}
&&
Y(t) = |g(t)|^{2}\,\left( \,
\int_{t_{i}}^{t}\,dt'\,\frac{\Gamma (t')S(t')}{|g(t')|^{2}} +
Y(t_{i})\,\frac{1}{|g(t_{i})|^{2}} \,\right) \,,
\label{integrated form of yield} 
\end{eqnarray}
where the source term $S(t)$ is given by
\begin{eqnarray}
&& \hspace*{0.5cm} 
\frac{1}{2\pi ^{2}\,T^{3}\,\Gamma (t)}\,\int_{0}^{\infty }\,dk\,k^{2}\,
\int_{\omega _{c}}^{\infty }\,d\omega \,
\nonumber 
\\ && 
\cdot \left( \,2\Re (g(t)h^{*}(\omega\,, \vec{k} \,, t)e^{-i\omega t}) 
+\Gamma (t)\,|h(\omega\,, \vec{k} \,, t)|^{2}\,\right)
\,\frac{\sigma (\omega \,, \vec{k})}{e^{\beta \omega } - 1} \,.
\end{eqnarray}
The stationary value of this source term $S(t)$
at $t\rightarrow \infty $ is
\begin{eqnarray}
&&
S_{\infty } = \frac{n_{\infty }}{T^{3}} \,, 
\end{eqnarray}
where $n_{\infty }$ is given by eq.(\ref{stationary n-density}).
Actually, this quantity is not stationary, since the temperature $T$ gradually
changes with the cosmological expansion.

When the rate $\Gamma _{k}$ does depend on the momentum $k$, the solution
for $Y$ cannot be given in a simple closed form such as 
eq.(\ref{integrated form of yield}).

\vspace{0.5cm} 
\section{Off-shell effect}
\vspace{0.5cm} 
\hspace*{0.5cm} 
The off-shell effect appears in two ways;
first, in the slower time dependence of relaxation ($|g| \propto 
t^{-\alpha - 1}$), and second, in a larger source term ($S
\propto T^{\alpha + 1}$ at small temperature $T$). 
Both are related to the threshold behavior of the spectral function, 
\( \:
\sigma \propto (\omega - \omega _{c})^{\alpha } \,.
\: \)

We found elsewhere \cite{jmy-97-1} that adding a power term $g_{1}(t)$ to the
pole term $g_{0}(t)$ is an excellent approximation for $g(t)$
in the entire time range,
unless one looks into the very short-time region of relaxation.
The power law period is represented by
\begin{eqnarray}
&&
g_{1}(t) = \frac{i\, c\Gamma (\alpha + 1)}{Q^{2}\,t^{\alpha + 1}}
\,e^{i(\omega _{c}t + \pi \alpha /2)} 
\,, \hspace{0.5cm} 
k_{1}(\omega \,, t) = -\,i\int_{t}^{\infty }\,dt'\,g_{1}(t')e^{-i\omega t'}
\,, 
\end{eqnarray}
where
\( \:
\sigma (\omega ) \approx c(\omega - \omega _{c})^{\alpha}
\: \)
near the threshold and
\( \:
Q = M - \omega _{c} \,.
\: \)
With this power behavior, the rate
\( \:
\Gamma (t) = -\,\frac{d}{dt}\ln |g(t)|^{2}
\: \)
is 
\( \:
\approx 2(\alpha + 1)/t \,.
\: \)
The final yield $Y = n/T^{3}$ is then of order,
\( \:
S_{\infty } \propto \frac{\Gamma }{M}\,(\frac{T}{M})^{\alpha + 1} \,.
\: \)
Thus, the yield does not decrease as rapidly as might have been expected
from the exponential decay law, but it decreases with a power,
\( \:
T^{\alpha + 1} \propto t^{-(\alpha + 1)/2} \,.
\: \)

A complication arises when one incorporates dependence on the particle
momentum.
This effect appears in two ways; first, in the time dilatation of
the lifetime $\omega _{k}/M$ and second, in the function $g(k \,, t)$.
The time dilatation effect is negligible if one only considers the
temperature range of $T < M$ (the mass of unstable particle).
We shall thus discuss the momentum dependence of the off-shell
contribution to $g$; $g_{1}(k\,, t)$.
In the $C_{1}$ contour integral of Fig.1 one has an approximate expression
of the form, with $\omega _{c} = k$,
\begin{eqnarray}
&&
g_{1}(k \,, t) \approx i\frac{e^{ikt}}{M^{2}}\,\int_{0}^{\infty }\,dy\,
\sigma (k + iy \,, k)\,e^{-yt}  \,.
\end{eqnarray}
We then use the spectral form (\ref{spectral form}) or its
low temerature approximation,
\begin{equation}
\sigma (\omega \,, k) \approx \frac{\Gamma }{2\pi }\,\frac{(\omega ^{2}
- k^{2})^{\alpha /2}}{M^{\alpha }} \,, 
\end{equation}
to get
\begin{eqnarray}
&&
g_{1}(k\,, t) \approx i(2i)^{\alpha /2}\Gamma (\frac{\alpha }{2} + 1)
\,\frac{\Gamma }{2\pi }\,\frac{k^{\alpha /2}\,e^{ikt}}{M^{\alpha + 2}\,
t^{\alpha /2 + 1}} \,.
\end{eqnarray}

An approximate evolution equation is summarized using the dimensionless
time variable,
\( \:
\tau \equiv \Gamma t \,, 
\: \)
and the time-temperature relation
\( \:
T/M = \sqrt{\eta /\tau } \,, 
\: \)
\begin{eqnarray}
&& \hspace*{-1.5cm}
\frac{dY}{d\tau } = -\,\int_{0}^{\infty }\,\frac{dk\,k^{2}}{2\pi ^{2}\,T^{3}}\,
\gamma (k\,, t)\,\left( \,f_{k}(t) - 
\,\int_{k}^{\infty }\,d\omega \,
\frac{\sigma (\omega \,, k)}{(\omega - \omega _{k} )^{2} 
+ (\Gamma /2)^{2}}\,\frac{1}{e^{\beta \omega } - 1} \,\right)\,,   
\\ &&
Y = \frac{1}{T^{3}}\,\int_{0}^{\infty }\,\frac{dk\,k^{2}}{2\pi ^{2}}\,
f_{k}(t) \,, 
\\ &&
\gamma (k\,, t) = -\,2\,\Re \,
\frac{d}{\Gamma \,dt}\ln (g_{0}(t) + g_{1}(k\,, t))
= -\,\frac{\frac{d}{\Gamma \,dt}|g_{0}(t) + g_{1}(k\,, t)|^{2}}
{|g_{0}(t) + g_{1}(k\,, t)|^{2}} \,.
\end{eqnarray}

Thus, the momentum dependence $g_{1}(k\,, t)$
is convoluted with other momentum dependent functions 
in the $\vec{k}$ integral.
In low temperatures the most important part of this momentum
dependence is in the spectral function $\sigma (\omega \,, k)$ 
which vanishes at the threshold,
$\omega _{c} = k$ for the massless daughter particles.
In some sample numerical computations that include the phase factor
in $g_{1}(k\,, t)$, 
we observe oscillatory behaviors around the transition time
from the pole to the power period that occurs at $\approx T_{{\rm eq}}$ 
given by (\ref{transition temperature}). 
In the rest of the time region
the yield $Y$ smoothly varies, and it is very well described by
using a simplified effective rate,
\begin{eqnarray}
&&
\gamma (k \,, t) \:\rightarrow  \:
\bar{\gamma }(k\,, t) = -\,\frac{\frac{d}{\Gamma \,dt}
\,(\,|g_{0}(t)|^{2} + |g_{1}(k\,, t)|^{2}\,)}
{|g_{0}(t)|^{2} + |g_{1}(k\,, t)|^{2}} \,.
\label{average rate} 
\end{eqnarray}
The oscillatory behavior in the transition region 
gets smoothed and approaches $\bar{\gamma }$
given above when one time-averages the rate and increases the resolution
time $\Delta t$ towards the lifetime $1/\Gamma $.
Moreover, the late time behavior is insensitive to whether one uses
the exact rate $\gamma (k\,, t)$ or the effective rate 
$\bar{\gamma }(k\,, t)$ above.
It is thus a reasonably good approximation to use the average rate
(\ref{average rate}).

In a still further simplification one may neglect the momentum dependence
in $|g_{1}(k \,, t)|^{2} \propto k^{\alpha }$ 
and replace the momentum by its average in low temperatures;
\( \:
k^{a} \rightarrow O[1] \times T^{a} \,.
\: \)
The relation $\overline{k^{\alpha }} \propto T^{\alpha }$ is consistent
with the late-time momentum distribution of the stationary occupation
number $f_{\infty }(\vec{k})$, as discussed 
in eq.(\ref{late-time distribution}).
We have checked that the following crude rate equation,
\begin{eqnarray}
&&
\frac{dY}{d\tau } = -\,\gamma \,(Y - Y_{0} - S_{0}) 
\,,
\\ &&
\gamma  = \frac{e^{-\Gamma t} + (\alpha + 2)B(\alpha )\,
(\frac{\Gamma }{M})^{\alpha + 4}\,(\frac{T}{M})^{\alpha }
\,(\Gamma t)^{-\alpha - 3}}
{e^{-\Gamma t} + B(\alpha )\,
(\frac{\Gamma }{M})^{\alpha + 4}\,(\frac{T}{M})^{\alpha }
\,(\Gamma t)^{-\alpha - 2}} \,, 
\\ &&
Y_{0} = \frac{1}{2\pi ^{2}\,T^{3}}\,
\int_{0}^{\infty }\,dk\,\frac{k^{2}}{e^{\sqrt{k^{2} + M^{2}}/T} - 1} \,, 
\\ &&
S_{0} = 
\frac{\zeta (\alpha + 4)\Gamma (\alpha + 4)
\Gamma (\frac{\alpha }{2} + 1)}
{16\pi ^{2}\,\sqrt{\pi }\,\Gamma (\frac{\alpha }{2} + \frac{5}{2})}\,
\frac{\Gamma }{M}(\frac{T}{M})^{\alpha + 1} \,,
\end{eqnarray}
is a reasonably good approximation.
We may use 
\( \:
B(\alpha ) = \frac{\Gamma (\frac{3}{2}\alpha + 3)}
{\Gamma (\frac{1}{2}\alpha  + 3)}
\,, 
\: \)
since 
\( \:
\overline{k^{\alpha }} = B(\alpha )\,T^{\alpha }
\: \)
for the late-time distribution function already discussed.
The source term is separated into the on-shell contribution $Y_{0}$
plus the off-shell contribution $S_{0}$.

There is a second temperature $T_{*}$ or time at which the off-shell effect
becomes conspicuous. This is the time when the two terms of $g(t)$
becomes equal;
\( \:
|g_{0}(t)|^{2} = \overline{|g_{1}(k\,, t)|^{2}} \,.
\: \)
Using the formula above, one gets in the $\Gamma \rightarrow 0$ limit
\begin{equation}
\frac{T_{*}}{M} \approx \sqrt{\frac{\eta }{(\alpha + 4)\,
\ln \frac{M}{\Gamma }}} \,.
\end{equation}
Unless $\eta $ is small (roughly $\eta < 1/\ln (\frac{M}{\Gamma })$), 
\( \:
T_{*} > T_{{\rm eq}} 
\: \)
usually. The transitional period then lasts for a while at
\( \:
T_{*} > T > T_{{\rm eq}}  \,.
\: \)
Thus, $T_{*}$ is the temperature at which the slower decrease
of the remnant becomes apparant, while $T_{{\rm eq}}$ is
the one at which the larger source term becomes visible.

The late time limit of solution to this equation 
can be analytically worked out.
Since the on-shell source term $Y_{0}$ is exponentially small at late times
when $T < T_{{\rm eq}}$,
\begin{equation}
\frac{dY}{d\tau } \approx -\,\frac{\alpha + 2}{\tau }\,
(\,Y - S_{0}\,) \,.
\end{equation}
This equation is readily solvable, and in the $\tau \rightarrow \infty $
limit
\begin{equation}
Y \:\rightarrow  \: \frac{2\alpha + 4}{\alpha + 3}\,
\frac{n_{\infty }}{T^{3}} =
\frac{2\alpha + 4}{\alpha + 3}
\,A(\alpha )\,\frac{\Gamma }{M}\,(\frac{T}{M})^{\alpha + 1}
\,.
\label{late time yield} 
\end{equation}
This asymptotic form becomes relevant, starting at $T = T_{{\rm eq}}$.

We show some numerical results in Fig.5 and compare with the simple analysis.
Besides $\alpha $, important parameters in the time evolution are
the rate $\Gamma /M$ and $\eta $ that appears in the time-temperature
relation (\ref{time-temperature relation}).
It is clearly seen that the yield $Y = n/T^{3}$ is accurately given by
the on-shell formula at high temperatures and by the power law
formula (\ref{late time yield}) at low temperatures, with a transition
region at 
\( \:
\frac{T}{M} \approx \; ({\rm several} \; \times 10)^{-1} \,.
\: \)

One may summarize this result, by saying that the time evolution
gives the yield $Y(t)$, starting from the stationary value 
$S_{\infty }(T)$ at high temperatures and ending at 
$\frac{2\alpha + 4}{\alpha + 3}S_{\infty }(T)$ at low temperatures.
The transient temperature is characterized by the temperature of order
$T_{{\rm eq}} - T_{*}$ and the yield below this temperature follows
\begin{equation}
Y \approx Y_{{\rm eq}}\,(\frac{T}{T_{{\rm eq}}})^{\alpha + 1} 
\,.
\end{equation}

Importance of the off-shell effect is measured by how close the
temperature Max$\left(T_{*} \,, T_{{\rm eq}}\right)$ for 
the onset of the off-shell effect is
to the temperarure scale for the decay. This temperature is estimated as
follows. One first defines $T_{d}$ by $\Gamma = H(T_{d})$, which
gives $T_{d}=\sqrt{\eta }\,M$.
We then discuss two cases separately. First, if $\eta \geq 1$, 
$T_{d} \geq M$ and the decay
and its inverse decay frequently occur at temperatures between
$T_{d}$ and the threshold $M$, maintaining the thermal abundance
$Y = Y_{0}$. At $T \leq M$ the inverse decay is
less frequent than the decay and there is a Boltzmann suppression
$e^{-M/T}$
until the temperature decreases down to $T_{*}$.
(In the case of $\eta \geq 1$, $T_{*} > T_{{\rm eq}}$ always.)
Thus, the importance of the off-shell effect is measured by how large
is the quantity,
\begin{equation}
\frac{T_{*}}{M} \approx \sqrt{\frac{\eta }{\ln \frac{M}{\Gamma }}} \,,
\end{equation}
if this number is less than unity.
It however often happens that $\eta \gg 1$ is huge and 
\( \:
T_{*} \gg M \geq T_{{\rm eq}} \,,
\: \)
as it occurs for instance in the top and the weak boson decay.
In this case a true measure of the off-shell importance is given by
how large the value of $\frac{T_{{\rm eq}}}{M} \approx 1/(\ln M/\Gamma )$
is.

On the other hand, if $\eta \ll 1$, then $T_{d} \ll M$, and
the decay does not occur until $T \leq  T_{d}$, much below the $Q$ value.
In this case the off-shell importance is determined by how large
\begin{equation}
\frac{{\rm Max}\;(T_{{\rm eq}}\,, T_{*})}{T_{d}} 
= {\rm Max}\;\left( \frac{1}{\sqrt{\eta }\,\ln \frac{M}{\Gamma }}
\,, \frac{1}{\sqrt{\ln \frac{M}{\Gamma }}}\right)
\,,
\end{equation}
is.

In these two cases of both large and small $\eta $, 
a large decay constant of
for example,
\( \:
\frac{\Gamma }{M} = 1 - 10^{-2} \,, 
\: \)
is expected to give a large off-shell effect irrespective of the
$\eta $ value.
This is physically reasonable since for a large coupling the narrow
width approximation is expected to break down.

We now turn to the scattering effect mentioned in Section 2.
We first give a crude estimate and later elaborate more quantitatively.
The loss of coherence due to the scattering off thermal particles
occurs in an inverse time scale of order
\begin{equation}
\sigma _{s} 
\approx n_{{\rm th}}(T)\,\frac{\overline{v\Sigma }}{2\pi } \,,
\end{equation}
where $\overline{v\Sigma }$ is the averaged cross section of the scattering 
interaction with the thermal medium.
This gives a new term to the spectral $\sigma $ of order
\begin{equation}
\sigma _{s} \approx 
\frac{4(2s + 1)\zeta (3)}{3\pi ^{2}}\,\frac{\alpha _{s}^{2}\,T^{3}}{M^{2}}
\,, 
\end{equation}
where $\alpha  _{s}$ is the strongest dimensionless coupling with
the medium and $2s + 1$ is the spin degrees of freedom. 
We took the Thomson type cross section $\propto \alpha _{s}^{2}/M^{2}$.
This is a small addition to the on-shell rate $\frac{\Gamma }{2\pi }$,
due to the extra coupling factor and the temperature suppression.
More important is its possible contribution 
to the late-time and the low-temperature region.
Its contribution to the occupation number in the stationary limit,
\begin{equation}
\delta f_{\infty } \approx \frac{1}{M^{2}}\,
\int_{\omega _{c}}^{\infty }\,d\omega \,\sigma _{s}\,e^{-\beta \omega }
\,, 
\end{equation}
gives
\begin{equation}
\delta f_{\infty } \approx 
\frac{4(2s + 1)\zeta (3)}{3\pi ^{2}}\,\alpha _{s}^{2}\,
(\frac{T}{M})^{4} \,.
\end{equation}

We have so far ignored the momentum dependence of the spectral
function $\sigma _{s}(\omega \,, \vec{q})$.
Indeed, the above estimate of the spectral function is valid only
near the mass shell.
The correct formula including the off-shell contribution is
\begin{eqnarray}
&&
\sigma _{s}(\omega \,, \vec{q}) = \frac{v\Sigma _{s}}{2\pi }\,
\int\,\frac{d^{3}k}{(2\pi )^{3}}\,e^{-\beta k}\,\theta 
(\omega + k - \sqrt{(\vec{k} + \vec{q})^{2} + M^{2}}) \,.
\end{eqnarray}
This gives to the stationary number density an extra off-shell term 
for the scattering effect,
\begin{equation}
\delta n_{\infty } = \frac{1}{M^{2}}\,\int\,\frac{d^{3}q}{(2\pi )^{3}}\,
\int\,d\omega \,e^{-\beta \omega }\,\sigma _{s}(\omega \,, \vec{q})
\approx \frac{2}{3}\,v\Sigma _{s}MT\,n_{{\rm th}}(T) \,.
\end{equation}
This formula is valid only when deviation from the thermal distribution
is small.
Compared to the on-shell value $n_{{\rm th}}$, this $\delta n_{\infty }$
is smaller by a factor $\alpha _{s}^{2}\,\frac{T}{M}$.
Thus, when the off-shell contribution of order,
\( \:
\frac{\Gamma }{M}\,(\frac{T}{M})^{\alpha + 1}\,T^{3} \,, 
\: \)
dominates over the on-shell value $n_{{\rm th}}$, 
the scattering effect to the source term $S$ may be ignored.

The scattering effect also gives a new late-time contribution;
\begin{equation}
\int\,\frac{d^{3}q}{(2\pi )^{3}}\,\delta g (\vec{q} \,,t) 
= \frac{1}{M^{2}}\,\int\,\frac{d^{3}q}{(2\pi )^{3}}\,
\int\,d\omega \,\sigma _{s}(\omega \,, \vec{q})\,
e^{i\omega t} \approx 
\frac{2(2s + 1)}{3\pi ^{2}}\,\alpha _{s}^{2}\,
\frac{T}{M^{2}\,t}\,n_{{\rm th}} \,, 
\end{equation}
to be compared to the previous off-shell contribution,
\begin{eqnarray}
&&
\int\,\frac{d^{3}q}{(2\pi )^{3}}\,|g_{1}(\vec{q} \,, t)| \approx 
\frac{\Gamma }{2\pi }\,(\frac{T}{M})^{\alpha /4 + \frac{3}{2}}\,
\frac{M^{2}}{(Mt)^{\alpha /2 + 1}} \,.
\label{off-shell integrated rate} 
\end{eqnarray}
Note that
\( \:
n_{{\rm th}}(T) < \frac{\Gamma }{M}\,(\frac{T}{M})^{\alpha + 1}\,T^{3}
\: \)
in the temperature region of our interest.
Thus, the scattering effect appears negligible, because the temperature
inequality usually obeyed,
\begin{equation}
\frac{T}{M} < O[\,(\frac{\Gamma }{\eta \alpha _{s}^{2}\,M})^{1/3}\,]
\,, 
\end{equation}
gives a smaller late-time contribution than 
eq.(\ref{off-shell integrated rate}).

\vspace{0.5cm} 
\section{Species evolution equation and baryogenesis}
\vspace{0.5cm} 
\hspace*{0.5cm} 
It is of great interest if one can predict consequences of
the modified abundance of unstable particles during their decay
that become permanently imprinted in the rest of the cosmic evolution.
In this respect it is important to clarify a possible change of the
thermal environment due to the decay.
The mere increase of the total number of decay products is hardly
recognizable in thermal medium.
One should examine a more detailed distribution of particle species
in the process of unstable particle decay.
As an important example of this class, we shall discuss baryogenesis
in a simplified model.

To this end we need to introduce several decay modes
distinguished by a flavor index $j$ of the continuous state
\( \:
b_{j}^{\dag }(\omega )|0\rangle \,.
\: \)
In the problem of baryogenesis we think of many channels of different
baryon numbers such as the diquark ($qq$) and the quark-lepton pair 
($\bar{q}\bar{l}$), considering the decay
of anti-$X$ boson along with the $X$-boson.
It is presumably better to consider channels of different $B - L$ such
as 
\( \:
N \rightarrow l\,\bar{H} \,, \; \bar{l}\,H 
\: \)
(where $N$ is a heavy right-handed Majorana neutrino, and $H$ is the Higgs
doublet), in view of that the baryon and the lepton numbers are 
redistributed at lower temperatures
by the baryon non-conserving electroweak process keeping $B - L$ unchanged.
\cite{dimopoulos-susskind}, \cite{krs87}, \cite{fukuyana86},
\cite{electroweak b-genesis review}

The basic assumption taken in estimating the environment change 
is that there exists 
baryon conserving strong interaction among environment particles such that
the kinetic equilibrium among them
is readily established, leading to the environment
particle distribution described by a thermal distribution function of
a finite chemical potential $\mu $ associated with the baryon number,
\begin{equation}
\langle b_{j}^{\dag }(\omega )b_{j}(\omega ) \rangle_{i}
= \frac{1}{e^{\beta (\omega - \alpha _{j}\mu )} - 1} \,.
\end{equation}
Here $\alpha _{j}$ is the baryon number of the species $j$.
In subsequent application to baryogenesis the limit of a small chemical
potential is relevant. Thus,
\begin{equation}
\langle b_{j}^{\dag }(\omega )b_{j}(\omega ) \rangle_{i}
\approx (1 - \alpha _{j}\,\mu \frac{d}{d\omega })\,
\frac{1}{e^{\beta \omega } - 1} \,.
\end{equation}

It is useful to note the conservation of particle number,
\begin{equation}
\frac{d}{dt}\left(\, c^{\dag }(t)c(t) +
\int_{\omega _{c}}^{\infty }\,d\omega \,\sum_{j}\,
b_{j}^{\dag }(\omega \,, t)b_{j}(\omega \,, t)  
\,\right) = 0 \,,
\end{equation}
which holds as an operator identity.
It is easy to confirm that if 
\begin{eqnarray}
&&
\langle b_{j}^{\dag }(\omega )b_{k}(\omega ') \rangle_{i}
\propto  \delta _{jk}\,\delta (\omega - \omega ') \,,
\nonumber 
\\ &&
\frac{d}{dt} \,\int_{\omega _{c}}^{\infty }\,d\omega \,
\sum_{j}\,\langle  b_{j}^{\dag }(\omega \,, t)b_{j}(\omega \,, t)\rangle
\rightarrow 0 \,, 
\end{eqnarray}
as $t \rightarrow \infty $, but 
a net baryon number may be generated, since
\begin{equation}
\frac{d}{dt}\,
\int_{\omega _{c}}^{\infty }\,d\omega \,\sum_{j}\,\alpha _{j}\,
\langle b_{j}^{\dag }(\omega \,, t)b_{j}(\omega\,, t )  \rangle \neq 0 \,.
\end{equation}

One can write down a set of time evolution equation of each species along 
with the one of the parent unstable particle.
These equations are not particularly illuminating.
We shall directly work out the situation relevant to baryogenesis.
In this problem the environment is gradually changed by the presence
of a small chemical potential associated with the baryon number.
We imagine that the bulk of thermalization processes not included
in the decay interaction conserves the baryon number, and
the time variation of the net baryon number is driven by
the asymmetry generated by the pair decay of $X$ and $\bar{X}$.
One thus expands in power series of the small chemical potential and 
identify the baryon number density of the thermal environment;
\begin{eqnarray}
n_{B} &=& \frac{1}{2\pi ^{2}}\,\int_{0}^{\infty }\,dk\,k^{2}\,
\sum_{j}\,\left( \,\frac{\alpha _{j}}{e^{\beta (\omega _{k} - \alpha _{j}\,
\mu ) }- 1} + (\,\alpha _{j} \rightarrow -\,\alpha _{j}\,)
\,\right)
\nonumber \\ 
&\approx& -\,\frac{\mu }{\pi ^{2}}\,\sum_{j}\,\alpha _{j}^{2}\,
\int_{0}^{\infty }\,dk\,k^{2}\,\frac{d}{d\omega _{k}}\frac{1}{e^{\beta 
\omega _{k}} - 1} \,.
\end{eqnarray}

Furthermore, we only consider for simplicity
the case of many decay modes whose rates differ only in the overall 
partial rate,
\( \:
\sigma _{j}(\omega ) = \gamma _{j}\,\sigma (\omega ) \,, 
\: \)
where the species independent $\sigma (\omega )$ is 
the total spectral function common to
$X$ and $\bar{X}$ due to the CPT theorem.
Thus 
\( \:
\sum_{j}\,\gamma _{j} = \sum_{j}\,\bar{\gamma }_{j} = 1 \,.
\: \)
This approximation is excellent if the decay products are much
lighter than the unstable particle, as it happens in the $X$ decay
into ordinary fermions.
The fundamental baryon asymmetry $\epsilon $
when a pair of $X$ and $\bar{X}$ bosons
decay arises via a combined effect of baryon non-conservation and
CP violation \cite{Sakharov} such that for some partial rates
\( \:
\gamma _{j} \neq \bar{\gamma }_{j} \,.
\: \)
It is given by
\begin{equation}
\epsilon \equiv 
\sum_{j}\,\alpha _{j}\,(\gamma _{j} - \bar{\gamma }_{j})
\,. \label{cp parameter} 
\end{equation}
We consider both the fermion-pair decay \cite{b-asymmetry for x} and the
boson-pair decay \cite{b-asymmetry via boson} of $X$ boson.
It is thus convenient to parametrize the spectral function as in 
(\ref{spectral form}),
\begin{equation}
\sigma (\omega \,, k) = \frac{\Gamma }{2\pi }\,\frac{M}{\omega _{k}}\,
\frac{(\omega ^{2} - k^{2})^{\alpha /2}}{M^{\alpha }} \,,
\end{equation}
where $\alpha = 2$ for the fermion-pair decay and $\alpha = 0$ for
the boson-pair decay.

To lowest order of the asymmetry $\epsilon $ and the chemical potential $\mu $,
the basic evolution equation is
\begin{eqnarray}
&& 
(\frac{d}{dt} + 3\frac{\dot{a}}{a})\,\,n_{B} = 
\int_{0}^{\infty }\,\frac{dk\,k^{2}}{2\pi ^{2}}\,
\left( \, \,\Gamma_{k} (t)\,
\left( \,
\frac{\epsilon }{2}(\,f_{X} + f_{\bar{X}}\,) 
+ \frac{\delta }{2}(\,f_{X} - f_{\bar{X}}\,) 
\,\right)  
\,\right. \nonumber 
\\ && \hspace*{0.5cm} 
-\,\epsilon \,
\int_{\omega _{c}}^{\infty }\,d\omega \,\sigma (\omega )\,
\left(\,\Gamma _{k}(t)|h(\omega \,, t)|^{2} +
2\Re (g(t)h^{*}(\omega \,, t)e^{-i\omega t})\,\right) f^{{\rm th}}(\omega )
\nonumber 
\\ && \hspace*{0.5cm} 
+ \,
\frac{\mu \delta^{2} }{2}\, 
\int_{\omega _{c}}^{\infty }\,d\omega \,\sigma (\omega )\,
\left(\,\Gamma _{k}(t)|h(\omega \,, t)|^{2} +
2\Re (g(t)h^{*}(\omega \,, t)e^{-i\omega t})\,\right)\,
\frac{df^{{\rm th}}(\omega )}{d\omega }
\nonumber 
\\ && \hspace*{0.5cm} 
\left.
+\,2\mu \,\left( \,\sum_{j}\alpha _{j}^{2}(\gamma _{j} + \bar{\gamma }_{j})
- \frac{\delta ^{2}}{2} \,\,\right)\,
\int_{\omega_{c}}^{\infty }\,d\omega \,\sigma (\omega )\,\Re h(\omega \,, t)
\,\frac{df^{{\rm th}}(\omega )}{d\omega }\,\right)   \,, 
\\ && 
(\frac{d}{dt} + 3\frac{\dot{a}}{a})\,(n_{X} + n_{\bar{X}})= 
\int_{0}^{\infty }\,\frac{dk\,k^{2}}{2\pi ^{2}}\,
\left( \, -\,\Gamma _{k}(t)\,(f_{X} + f_{\bar{X}}) 
\right. \nonumber 
\\ && 
\left. + \,
2\,\int_{\omega _{c}}^{\infty }\,d\omega \,
\left(\,\Gamma _{k}(t)|h(\omega \,, t)|^{2} +
2\Re (g(t)h^{*}(\omega \,, t)e^{-i\omega t})\,\right)\,\sigma (\omega )\,
f^{{\rm th}}(\omega ) 
\,\right) \,, 
\\ && 
(\frac{d}{dt} + 3\frac{\dot{a}}{a})\,(n_{X} - n_{\bar{X}})= 
\int_{0}^{\infty }\,\frac{dk\,k^{2}}{2\pi ^{2}}\,
\left( \,
-\,\Gamma _{k}(t)\,(f_{X} - f_{\bar{X}}) \, 
\,\right.
\nonumber 
\\ && \hspace*{-1cm}
\left. \,- \mu\, \delta \,
\int_{\omega _{c}}^{\infty }\,d\omega \,(\,
\sigma (\omega )\,
\left(\,\Gamma _{k}(t)|h(\omega \,, t)|^{2} +
2\Re (g(t)h^{*}(\omega \,, t)e^{-i\omega t})\,\right)\,
\frac{df^{{\rm th}}(\omega )}{d\omega }
\,\right)
\,. \label{time evolution eq for b} 
\end{eqnarray}
Here 
\begin{equation}
f^{{\rm th}}(\omega ) = \frac{1}{e^{\beta \omega } - 1}
\end{equation}
is the thermal occupation number for the zero chemical potential and
\begin{equation}
\delta = \sum_{j}\,\alpha _{j}(\gamma _{j} + \bar{\gamma }_{j}) \,.
\end{equation}

We first consider the pole dominance approximation.
In this approximation a simple relation,
\begin{equation}
\Re h_{p} = \Re \left( \,-\,\frac{\dot{g}}{g}|h|^{2} 
+ gh^{*}e^{-i\omega t}\,\right)_{p} \,, 
\end{equation}
holds.
Using the relation of the baryon number density and the chemical potential,
\begin{equation}
n_{B} = \frac{1}{3}\,
\sum_{j}\alpha _{j}^{2}\,\mu T^{2}\,
\,, 
\end{equation}
the relevant evolution equation is 
\begin{eqnarray}
&& 
(\,\frac{d}{dt} + 3\frac{\dot{a}}{a}\,)\,n_{B} =
\epsilon \,\Gamma \, \left(\,n_{+} - n_{0}^{{\rm th}}\,\right) 
+ \Gamma \delta \,n_{-} - \frac{K}{2}\,\Gamma \,
\frac{n_{0}^{{\rm th}}}{T^{3}}\,n_{B}
 \,,
\\ && 
(\,\frac{d}{dt} + 3\frac{\dot{a}}{a}\,)\,n_{+}
= -\,
\Gamma \,\left(\,n_{+} - n_{0}^{{\rm th}}
\right) 
\,, \label{average number eq} 
\\ &&
(\,\frac{d}{dt} + 3\frac{\dot{a}}{a}\,)\,n_{-}
= -\,\Gamma n_{-} + \frac{\tilde{\delta} \Gamma }{2}\,
\frac{n_{0}^{{\rm th}}}{T^{3}}\,n_{B} \,, 
\\ &&
n_{0}^{{\rm th}} = (\frac{MT}{2\pi })^{3/2}\,e^{-\,M/T} \,, 
\\ &&
n_{+} = \int_{0}^{\infty }\,\frac{dk\,k^{2}}{4\pi ^{2}}\,
(f_{X} + f_{\bar{X}})
\,, \hspace{0.5cm} 
n_{-} = \int_{0}^{\infty }\,\frac{dk\,k^{2}}{4\pi ^{2}}\,
(f_{X} - f_{\bar{X}}) \,, 
\\ && 
K = 
\frac{6\,\sum_{j}\,\alpha _{j}^{2}(\gamma _{j} + \bar{\gamma }_{j})}
{\sum_{j}\,\alpha _{j}^{2}} > 0 
\,, \hspace{0.5cm} 
\tilde{\delta } = \frac{3\delta }{\sum_{j}\,\alpha _{j}^{2}}
\,.
\end{eqnarray}
We simplified the equation, using the condition of very low temperature,
\( \:
T \ll M \,.
\: \)
In the numerical computation below, we allow the temperature region,
\( \:
T \approx M \,.
\: \)

Since the bulk of the cosmic medium is in thermal equilibrium,
the usual \\ 
temperature-time relation (\ref{temperature-time rel}) holds.
One can then integrate the equation for the average number density
(\ref{average number eq}), which gives a source term $Y_{S}$
to the rest of equations;
\begin{eqnarray}
&&
Y_{S}(\tau ) \equiv \eta ^{3/4}\,
\frac{n_{X} - n_{0}^{{\rm th}}}
{T^{3}} \nonumber 
\\ && \hspace*{-1.5cm}
\,=
(2\pi )^{-3/2}\,\left( \,
e^{-\,\tau }\,\int_{\tau _{i}}^{\tau }\,dx\,x^{3/4}\,e^{x - \sqrt{x/\eta }}
- \tau ^{3/4}\,e^{-\sqrt{\tau /\eta }}\,\right) 
+ \eta ^{3/4}\,e^{-(\tau - \tau _{i})}\,(\frac{n_{+}}{T^{3}})_{i} 
\,, 
\\ &&
\frac{dY_{B}}{d\tau } = -\,\frac{K}{2}\,Y_{0}\,Y_{B} 
+ \delta Y_{-} + \epsilon\, \eta ^{-3/4}\,Y_{S} \,, 
\\ &&
\frac{dY_{-}}{d\tau } = -\,Y_{-} + \frac{\tilde{\delta }}{2}\,
Y_{0}\,Y_{B} \,, 
\\ && \hspace*{1cm}
Y_{B} \equiv \frac{n_{B}}{T^{3}} \,, \hspace{0.5cm} 
Y_{-} \equiv \frac{n_{-}}{T^{3}} \,,
\hspace{0.5cm} 
Y_{0} \equiv \frac{n_{0}^{{\rm th}}}{T^{3}} \,.
\end{eqnarray}
Solutions to this approximation will be compared to a more precise
numerical result.

A comparison with previous works \cite{kolb-wolfram},\cite{fry-olive-turner}
reveals some differences. 
In our treatment two-body processes 
\( \:
b^{\dag }_{j}(\omega )|0\rangle 
\leftrightarrow b^{\dag }_{k}(\omega )|0\rangle  
\: \)
are not included. 
(In the terminology of \cite{kolb-wolfram} this corresponds to the
off-shell two-body contribution and their on-shell two-body
term is automatically included in our approach, too.)
This is a higher order effect, hence was neglected in the present work.
The major difference absent in the past work and included here
is however the off-shell effect, as will be discussed shortly.

An improved, but a still simple approximation that incorporates
the power law period of decay is to add the two contributions in
the exponential and the power law periods incoherently, 
ignoring the interference between the pole and the power terms.
This introduces a time, or temperature dependence of the rate
$\Gamma (t)$ that includes the off-shell effect.

Another simplification is necessary for a practical computation
of momentum integrals containing the momentum dependent 
effective rate and the time dilatation factor.
It is not difficult to partially include the momentum dependence
in quantities such as
\begin{equation}
\int_{0}^{\infty }\,\frac{dk\,k^{2}}{2\pi ^{2}}\,\bar{\gamma }(k\,, t)\,
\int_{k}^{\infty }\,d\omega \,\frac{\sigma (\omega \,, k)}
{(\omega - \omega _{k})^{2} + (\pi \,\sigma (\omega \,, k))^{2}}\,
\frac{1}{e^{\beta \omega } - 1} \,.
\end{equation}
Indeed we did perform these momentum integrations in our numerical analysis.
But unless the integro-differential equation for the
occupation number, instead of the ordinary differential equation for
the number density, is directly solved, it is difficult to deal with
the momentum dependence in the convolution integral containing
$f_{X}$ and $f_{\bar{X}}$ in eqs.(\ref{time evolution eq for b}).
This is because these distributions are unknown before we know the
complete answer to the problem, which is a formidable task.
Thus, we made a replacement of the momentum factor by a temperature factor 
in those terms containing $f_{X}$ and $f_{\bar{X}}$;
\begin{equation}
k^{\alpha } \:\rightarrow  \: 
\frac{\Gamma (\frac{3}{2}\alpha + 3)}{\Gamma (\frac{1}{2}\alpha  + 3)}
\,T^{\alpha }
\,,
\end{equation}
using the momentum distribution function relevant at late times.

It turns out that the rest of the detailed momentum dependence 
integrated numerically is 
not crucial; its variation only changes the transient time dependence around
the temperature
$T_{*}$ and quantities in the rest of the time region are insensitive
to this momentum dependence.
We therefore write down a simplified rate equation replacing the
momentum dependence by its average above.
In terms of the net number $Y_{\pm } \equiv n_{\pm }/T^{3}$ and
the rescaled time $\tau \equiv \Gamma t$,
\begin{eqnarray}
&& 
\left\{
\begin{array}{l}
\frac{dY_{+}}{d\tau } = -\,\gamma \,
\left( Y_{+} - Y_{0} - S_{0} \right)
\,,   
\\
\frac{dY_{-}}{d\tau } = -\,\gamma \,
\left( \,
Y_{-} - \frac{\tilde{\delta }}{2}\,(Y_{1} + S_{1})\,Y_{B} 
\,\right)
\,,
\\
\frac{dY_{B}}{d\tau } = \gamma \,
\left( \,\epsilon 
Y_{+} + \delta \,Y_{-} - \epsilon (Y_{0} + S_{0}) 
- \frac{\delta \tilde{\delta }}{2}\,(Y_{1} + S_{1})\,Y_{B}\,\right)
- c \,(Y_{1} + S_{2})\,Y_{B} 
\,, 
\end{array}
\right. 
\nonumber \\ &&
\\ && 
\gamma  = \frac{e^{-\Gamma t} + (\alpha + 2)B(\alpha )\,
(\frac{\Gamma }{M})^{\alpha + 4}\,(\frac{T}{M})^{\alpha }
\,(\Gamma t)^{-\alpha - 3}}
{e^{-\Gamma t} + B(\alpha )\,
(\frac{\Gamma }{M})^{\alpha + 4}\,(\frac{T}{M})^{\alpha }
\,(\Gamma t)^{-\alpha - 2}} \,, \hspace{0.5cm} 
B(\alpha ) = \frac{\Gamma (\frac{3}{2}\alpha + 3)}
{\Gamma (\frac{1}{2}\alpha + 3)} \,, 
\\ &&
S_{0} = 
\frac{\zeta (\alpha + 4)\Gamma (\alpha + 4)
\Gamma (\frac{\alpha }{2} + 1)}
{16\pi ^{2}\,\sqrt{\pi }\,\Gamma (\frac{\alpha }{2} + \frac{5}{2})}
\,\frac{\Gamma }{M}\,(\frac{T}{M})^{\alpha + 1} \,,
\\ && \hspace*{0.5cm} 
c = 
\frac{3\,(\,\sum_{j}\,\alpha _{j}^{2}(\gamma _{j} + \bar{\gamma }_{j})
- \frac{\delta ^{2}}{2}\,)}{\sum_{j}\,\alpha _{j}^{2}} > 0
\,, 
\label{constant c} 
\\ &&
S_{1} = S_{1}(\alpha ) = 
\frac{\zeta (\alpha + 3)}{8\pi ^{2}\sqrt{\pi }}\,
\frac{\Gamma (\frac{\alpha }{2} + 1)\Gamma (\alpha + 3)}
{\Gamma (\frac{\alpha }{2} + \frac{3}{2})}
\,\frac{\Gamma }{M}\,(\frac{T}{M})^{\alpha + 1}
\,, 
\\ &&
S_{2} = S_{1}(2\alpha ) =
\frac{\zeta (2\alpha + 3)}{8\pi ^{2}\sqrt{\pi }}\,
\frac{\Gamma (\alpha  + 1)\Gamma (2\alpha + 3)}
{\Gamma (\alpha + \frac{3}{2})}
\,\frac{\Gamma }{M}(\frac{T}{M})^{2\alpha + 1} \,, 
\\ &&
Y_{1} = \frac{1}{2\pi ^{2}\,T^{2}}\,
\int_{0}^{\infty }\,dk\,\frac{2k^{2} + M^{2}}{\sqrt{k^{2} + M^{2}}}\,
\frac{1}{e^{\sqrt{k^{2} + M^{2}}/T} - 1} 
\,.
\end{eqnarray}
The low temperature approximation was not assumed here for
$f^{{\rm th}}(\omega )$, hence
\begin{equation}
Y_{0} = \frac{1}{2\pi ^{2}\,T^{3}}\,
\int_{0}^{\infty }\,dk\,\frac{k^{2}}{e^{\sqrt{k^{2} + M^{2}}/T} - 1} \,.
\end{equation}

It can be readily proved by a rescaling argument that 
both $Y_{-}$ and the baryon asymmetry $Y_{B}$ is in direct proportion to
the fundamental CP parameter $\epsilon $.
We assume that $\alpha \leq 2$ as required for any renormalizable
decay interaction.

Some results of numerical integration of the time evolution equation
are presented in Fig.6 and Fig.7.
The time evolution for $\alpha = 0$ and $\alpha = 2$ is evidently
different, as seen in Fig.6.
Notably, the final $Y_{B}$ vanishes for $\alpha = 0$.
This difference will be understandable analytically, as will be discussed
shortly.
The final amount of the baryon to the photon ratio, approximately
given by $Y_{B} = \frac{n_{B}}{T^{3}}$, is shown as a function of $\eta $
in Fig.8,
the ratio of the decay rate to the Hubble rate at $T = M$.
In all these computations of Fig.6 $-$ 8 the initial $X$ abundance
was assumed to be the thermal value.
In Fig.9 we show how the time evolution is affected by the initial
condition, taking grossly different $X$ boson abundance from
the thermal value.
In Fig.6 $-$ 9 we used for the parameters given by the underlying theory,
\( \:
\delta = \frac{1}{3} \,, \; \tilde{\delta } = \frac{9}{5} \,, \;
c = 6 \,, 
\: \)
chosen to be consistent with the CPT constraint.

The asymptotic behavior of solutions to this rate equation differs, depending
on whether $\alpha > \frac{1}{2}$ or $\alpha \leq \frac{1}{2}$.
First, let us assume that the baryon asymmetry $Y_{B}$ approaches
a constant (including 0) asymptotically as $t \rightarrow \infty $.
The first and the second equation for $Y_{\pm }$ in the above set then gives
the asymptotic solution;
\begin{eqnarray}
&&
Y_{+} \:\rightarrow  \: \frac{2\alpha + 4}{\alpha + 3}\,
\frac{n_{\infty }}{T^{3}}
\,,
\\ &&
Y_{-} \:\rightarrow  \: \frac{3(\alpha + 2)}{\alpha + 3}\,\frac{\delta Y_{B}}
{\sum_{j}\,\alpha _{j}^{2}}\,S_{1} \,.
\end{eqnarray}
Both $Y_{\pm } \propto \frac{\Gamma }{M}(\frac{T}{M})^{\alpha + 1}
\propto t^{-(\alpha + 1)/2}$.
The asymptotic equation for $Y_{B}$ differs, depending on whether
$\alpha \neq 2$ or $\alpha = 2$. We shall first discuss the simpler case
of $\alpha \neq 2$.
The equation for the asymmetry in this case is 
\begin{equation}
\frac{dY_{B}}{d\tau } \approx -\,c\,S_{2}\,Y_{B} \,,
\end{equation}
where $c$ is given by eq.(\ref{constant c}).
Since 
\( \:
S_{2} \propto \tau ^{-\,(2\alpha + 1)/2} \,, 
\: \)
\begin{eqnarray}
Y_{B} \approx {\rm const.}\; \exp [-\,c\,\int_{\tau_{0} }^{\tau }\,dx\,
S_{2}(x)] 
\label{asymptotic b-asym} 
\end{eqnarray}
approaches a constant for $\alpha > \frac{1}{2}$.
Thus, as $t \rightarrow \infty $, $Y_{B}$ approaches a constant from
above, its rate to the asymptote being of order
\( \:
O[t^{-(2\alpha - 1)/2}] \,.
\: \)
When $\alpha = 2$, the asymptotic equation for $Y_{B}$ is more complicated,
but the asymptotic behavior of the asymmetry is identical to the case for
$2 > \alpha > \frac{1}{2}$.

On the other hand, for $\alpha \leq \frac{1}{2}$ the integral in
the exponent of eq.(\ref{asymptotic b-asym})
is divergent as $\tau \rightarrow \infty $, hence $Y_{B} \rightarrow 0$.
The asymptotic behavior is as follows:
\( \:
Y_{B} = O[e^{- c\,\Gamma \,t^{1/2 - \alpha }}]
\: \)
with $c > 0$ for $\alpha < \frac{1}{2}$, and
\( \:
Y_{B} = O[t^{- c\,\Gamma\, }]
\: \)
for $\alpha = \frac{1}{2}$.
In superrenormalizable models of the boson-pair decay $\alpha = 0$,
and this case gives a vanishing asymptotic baryon asymmetry, unless
bosons in the decay product quickly decay further
into ordinary quarks and leptons.

The amount of the generated baryon asymmetry in the large $\eta $ limit 
is of interest, because in the usual treatment of the out-of equilibrium
condition a very large mass is demanded for the $X$ boson 
due to the on-shell kinematics \cite{b-asymmetry review}:
it requires that the ratio of the decay rate to the Hubble rate at
the temperature $T = M$, precisely $\eta $, should not be too large
in order to get a sizable $X$ abundance for the baryon generation.
More precisely, the on-shell Boltzmann approach gives the asymmetry that
depends on $\eta $ like $\propto  \eta ^{-1.2}$ for a large $\eta $
\cite{kolb-wolfram}, \cite{fry-olive-turner}.
In Fig.8 we observe that the off-shell effect gives a less decreasing
behavior as $\eta $ increases.
It would be of some interest if one can work out the infinite
$\eta $ limit analytically.

According to the recent analysis of the reheating problem after inflation
it is quite possible \cite{x production at preheating} 
that the heavy $X$ boson has been created right after
the explosive decay \cite{explosive inflaton decay} 
of the inflaton oscillation and prior to the thermalization process. 
It is thus of considerable interest to examine the baryogenesis
without assuming the thermal abundance of the heavy boson initially.
The non-thermal initial condition has been taken in Fig.8 and for a large
enough $\eta $ the final baryon asymmetry is seen insensitive to
the initial condition.

\vspace{2cm}
In summary, we derived the cosmological time evolution equation for
the abundance of unstable particles, including the off-shell effect
not taken into account in the Boltzmann approach.
Application to the baryogenesis problem shows that the out-of equilibrium
condition based on the on-shell kinematics is changed.

\vspace{2cm}
\begin{center}
{\bf Acknowledgment}
\end{center}

This work has been supported in part by the Grand-in-Aid for Science
Research from the Ministry of Education, Science and Culture of Japan,
No. 08640341.
The work of the author (I.J.) is supported by a fellowship of
the Japan Society of the Promotion of Science.

\newpage

\begin{Large}
\begin{center}
{\bf Figure caption}
\end{center}
\end{Large}

\vspace{0.5cm} 
\hspace*{-0.5cm}
{\bf Fig.1}

Contour for the $\omega $ integral that separates the on-shell contribution
($C_{0}$) from the off-shell one ($C_{1}$). The dashed parts are in the
second Riemann sheet continued via the cut starting from the threshold
$\omega _{c}$.

\vspace{0.5cm} 
\hspace*{-0.5cm}
{\bf Fig.2}

Momentum distribution for a temperature and a decay rate,
\( \:
T = 0.07\,M \,, \; \Gamma = 0.01\,M \,.
\: \)
A comparison is made between the on-shell distribution function,
$e^{-\,\sqrt{k^{2} + M^{2}}/T}$ appropriate at low temperatures, and
the off-shell distribution of $\alpha = 0$ and $\alpha = 2$.
Two of these distributions should be multiplied by 0.01 to get the correct
values.

\vspace{0.5cm} 
\hspace*{-0.5cm}
{\bf Fig.3}

Stationary number density given by eq.(\ref{stationary n-density})
which is divided by temperature$^{3}$
for two values of the decay rate,
\( \:
\frac{\Gamma }{M} = 0.1 \,, \; 0.01 \,.
\: \)
The dotted lines are calculated using the approximate formula,
eq.(\ref{approximate stationary f}), while the broken line is the
on-shell contribution alone, the first term in this equation.

\vspace{0.5cm} 
\hspace*{-0.5cm}
{\bf Fig.4}

The equal-temperature at which the on-shell and the off-shell contributions
become equal is shown for two values of 
\( \:
\alpha = 0 \,, \; 2 \,.
\: \)
The dashed and the dotted lines are result of the approximate formula,
eq.(\ref{transition temperature}).

\vspace{0.5cm} 
\hspace*{-0.5cm}
{\bf Fig.5}

Time evolution of the yield $Y = \frac{n}{T^{3}}$ for different
values of $\eta $; 0.1 (Fig.5a) and 1 (Fig.5b) and for different
decay rates,
\( \:
\frac{\Gamma }{M} = 0.1 \,, \; 0.01 \,.
\: \)
For comparison the on-shell evolution (the broken line) and the late time
form, eq.(\ref{late time yield}) (the dotted line) are shown.

\vspace{0.5cm} 
\hspace*{-0.5cm}
{\bf Fig.6}

Comparison of the time evolving baryon asymmetry.
The case of $\alpha = 0$ shown by the solid line and enlarged in the inlet
gives the vanishing
value for the final asymmetry, unlike the $\alpha = 2$ case shown by
the dotted line. 

\newpage
\hspace*{-0.5cm}
{\bf Fig.7}

Time evolution of the baryon asymmetry.
Two cases of different decay rates,
\( \:
\frac{\Gamma }{M} = 0.1 \,, 0.01 \,, 
\: \) 
are compared to the evolution given by
the on-shell contribution alone (the broken line).
In the inlet detailed behaviors are stressed.

\vspace{0.5cm} 
\hspace*{-0.5cm}
{\bf Fig.8}

Final amount of the baryon asymmetry plotted against $\eta $
(= decay rate/Hubble rate at $T = M$).
For comparison the result based on the on-shell contribution alone
is shown by the dashed line.
Those marked by open boxes and circles are results for smaller decay
rates $\frac{\Gamma }{M}$.

\vspace{0.5cm} 
\hspace*{-0.5cm}
{\bf Fig.9}

Dependence of the asymmetry on the initial $X$ abundance for
different $\eta $ values; 0.1 (Fig.9a) and 10 (Fig.9b).
Result for the initially thermal abundance given by the solid line
is compared to those of 10 times the thermal value (the dotted line) and
the zero abundance (the dashed line).


\vspace{1cm}
\begin{thebibliography}{99}

\bibitem{kolb-turner book}
For a review, see E.W. Kolb and M.S. Turner, {\em The Early Universe}
(Addison Wesley, California, 1990).


\bibitem{jmy-96}
I. Joichi, Sh. Matsumoto and M. Yoshimura,
{\em Quantum Dissipation and Decay in Medium}, TU/96/510 and
hep-th/9609223, and {\sl Phys.\ Rev. }{\bf A57} No.2 in press; \\
{\sl Prog.\ Theor.\ Phys.\ }{\bf 98}, 9(1997), and cond-mat/9612235.

\bibitem{gut b-asymmetry}
M. Yoshimura, Phys.Rev.Lett.{\bf 41}, 281(1978);
Phys.Lett.{\bf 88B}, 294(1979).\\
S. Dimopoulos and L. Susskind, Phys.Rev.{\bf D18}, 4500(1978).\\
D. Toussaint, S.B. Treiman, F. Wilczek and A. Zee, Phys.Rev.{\bf
D19}, 1036(1979).\\
S. Weinberg, Phys.Rev.Lett.{\bf 42},850(1979).


\bibitem{b-asymmetry review}
For reviews, see \\
M. Yoshimura, {\em Cosmological Baryon Production and Related
Topics},
in Proceedings of the 4th Kyoto Summer Institute on
{\em Grand Unified Theories and Related Topics},
(World Scientific, Singapore, 1981);\\
{\em Baryogenesis and thermal history after
inflation}, 
{\em Journal of the Korean Physical Society}\,{\bf 29}, S236(1996)
and hep-th/9605246.\\
E.W. Kolb and M.S. Turner, {\sl Annu.\ Rev.\ Nucl.\ Part.\ Sci.\ }
{\bf 23}, 645(1983).


\bibitem{jmy-97-1} 

I. Joichi, Sh. Matsumoto and M. Yoshimura,
{\em Time Evolution of Unstable Particle Decay Seen with Finite Resolution},
TU/97/524 and hep-ph/9711235.


\bibitem{dimopoulos-susskind}
S. Dimopoulos and L. Susskind, Phys.Rev.{\bf D18}, 4500(1978).

\bibitem{krs87} 
V.A. Kuzmin, V.A. Rubakov and M.E. Shaposhnikov, 
{\sl Phys.\ Lett.\ }{\bf 155B}, 36(1985).

\bibitem{fukuyana86} 
M. Fukugita and T. Yanagida, {\sl Phys.\ Lett.\ }{\bf 174B}, 45(1986).

\bibitem{electroweak b-genesis review}
For a review of electroweak baryogenesis, 
A.G. Cohen, D.B. Kaplan and A.E. Nelson,
{\sl Annu.\ Rev.\ Nucl.\ Part.\ Sci.\ }{\bf 43}, 27(1993).


\bibitem{Sakharov}
A.D. Sakharov, {\sl JETP Lett.\ }{\bf 5}, 24(1967).

\bibitem{b-asymmetry for x}
D.V. Nanopoulos and S. Weinberg, {\sl Phys.\ Rev. }{\bf D20}, 2484(1979).\\
S. Barr, G. Segre and H.A. Weldon, {\sl Phys.\ Rev. }{\bf D20}, 2494(1979).\\
T. Yanagida and M. Yoshimura, {\sl Nucl.\ Phys. }{\bf B168}, 534(1980).\\
A. Yildiz and P. Cox, {\sl Phys.\ Rev. }{\bf D21}, 906(1980).

\bibitem{b-asymmetry via boson}
S. Barr, G. Segre and H.A. Weldon, {\sl Phys.\ Rev. }{\bf D20}, 2494(1979).

\bibitem{kolb-wolfram}
E.W. Kolb and S. Wolfram, {\sl Phys.\ Lett.\ }{\bf 91B}, 217(1980);
{\sl Nucl.\ Phys.\ }{\bf B172}, 224(1980).

\bibitem{fry-olive-turner}
J.N. Fry, K.A. Olive and M.S. Turner, {\sl Phys.\ Rev.\ Lett.\ }{\bf 45},
2074(1980);
{\sl Phys.\ Rev. }{\bf D22}, 2953, 2977(1980).

\bibitem{x production at preheating}
H. Fujisaki, K. Kumekawa, M. Yamaguchi and M. Yoshimura,
{\sl Phys.\ Rev. }{\bf D54}, 2494(1996);
M. Yoshimura,
"Baryogenesis and Thermal History after Inflation",
{\em Journal of the Korean Physical Society}\,{\bf 29}, S236(1996) and
hep-ph/9605246.\\
E.W. Kolb, A. Linde, and A. Riotto,
{\sl Phys.\ Rev.\ Lett.\ }{\bf 77}, 4290(1996).

\bibitem{explosive inflaton decay}
L. Kofman, A. Linde, and A.A. Starobinsky,
{\sl Phys.\ Rev.\ Lett.\ }{\bf 73}, 3195(1994).

\end{thebibliography}
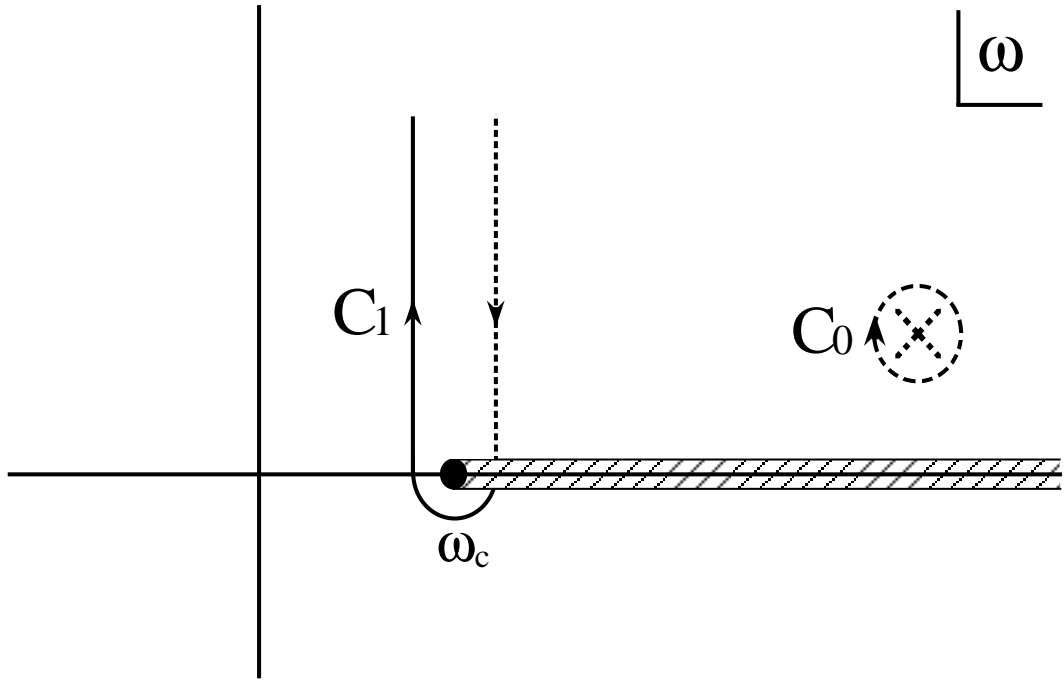
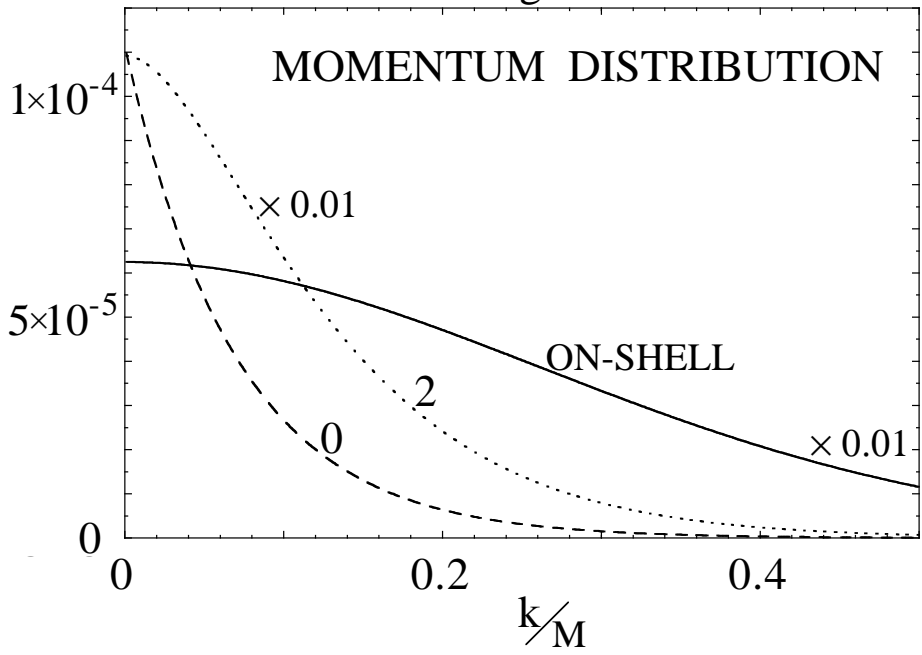
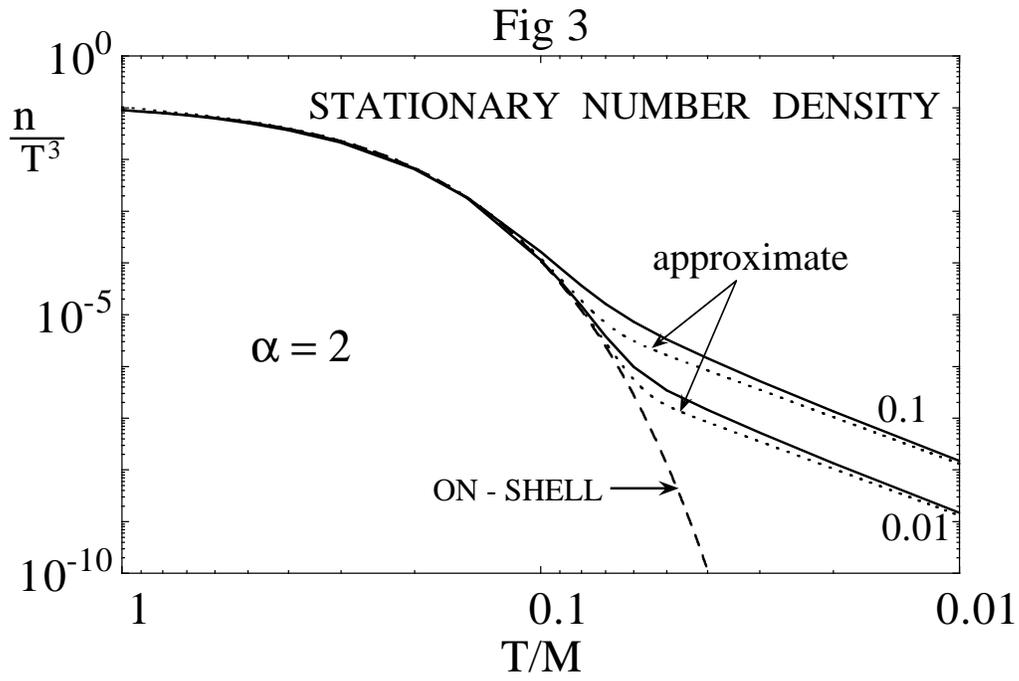
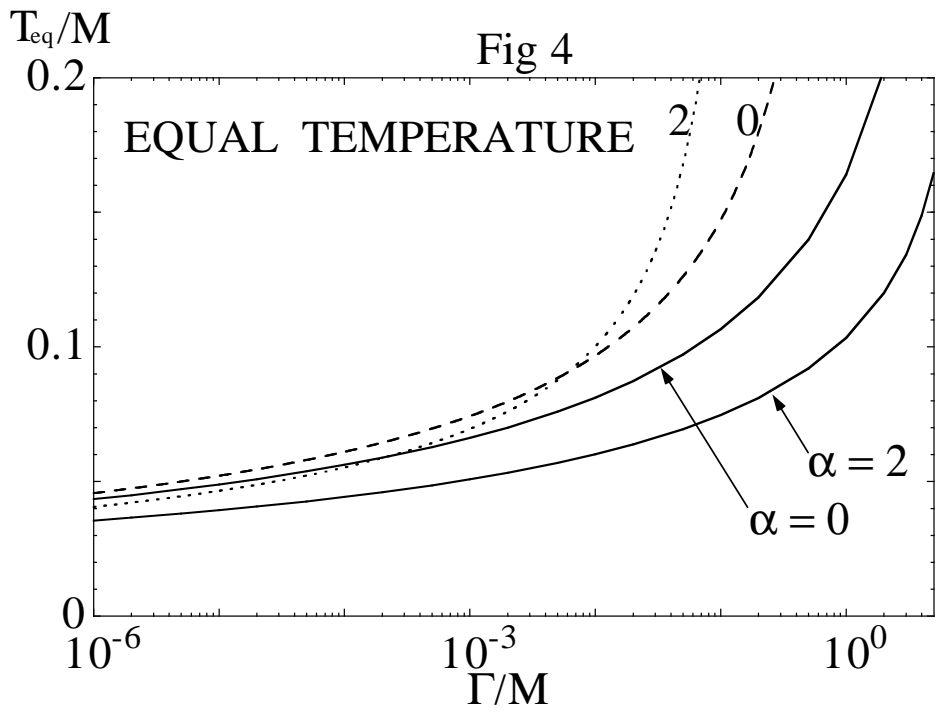
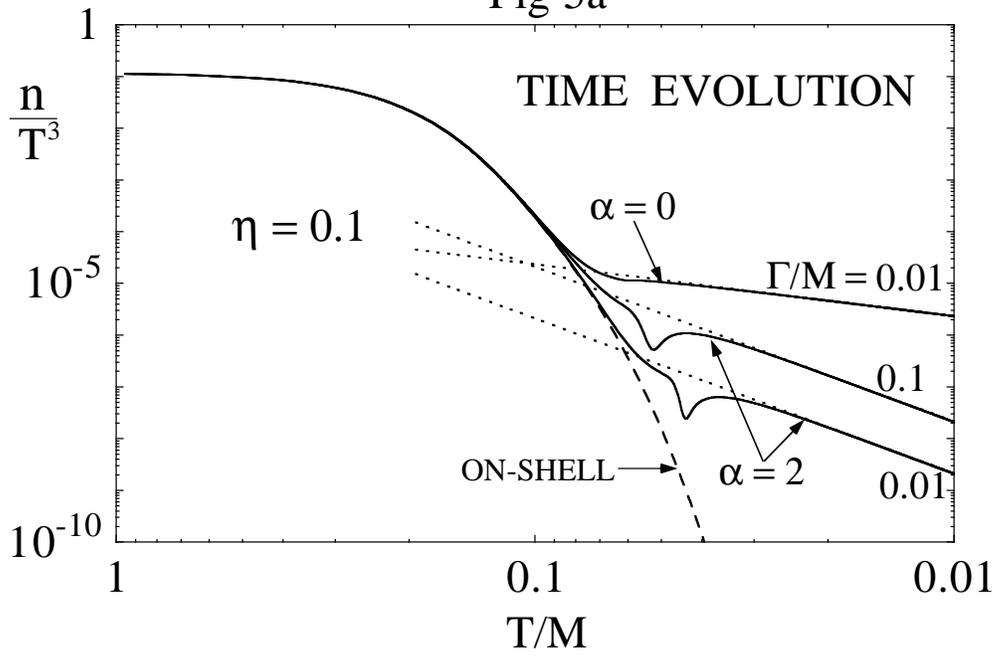
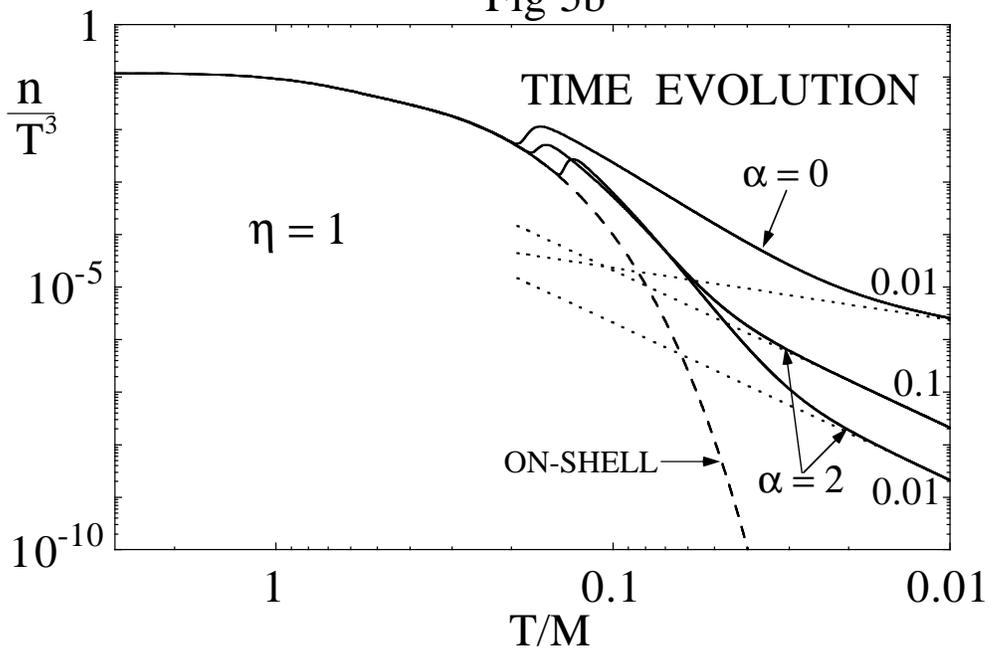
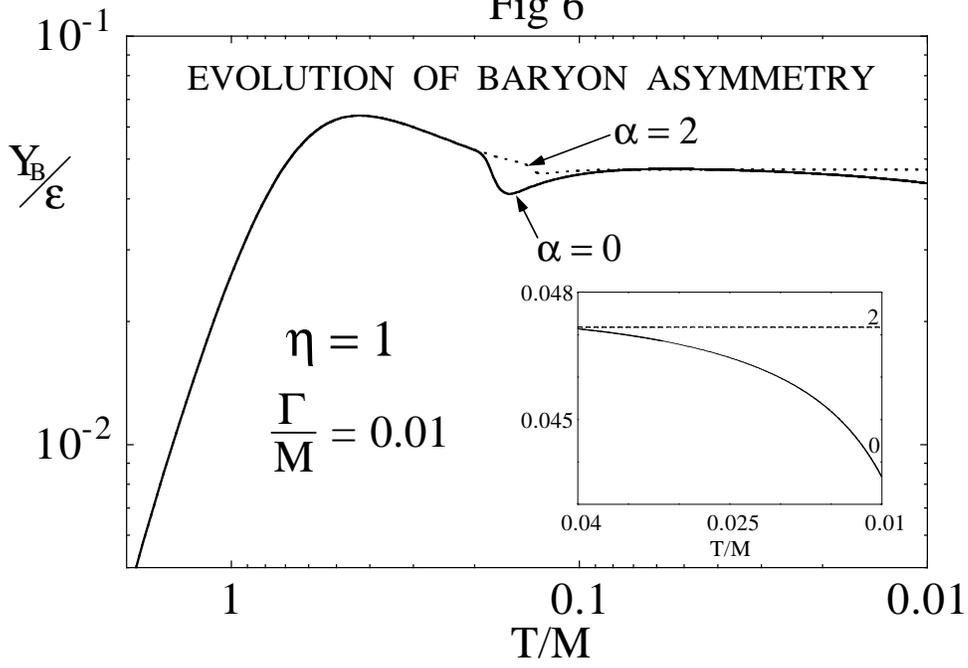
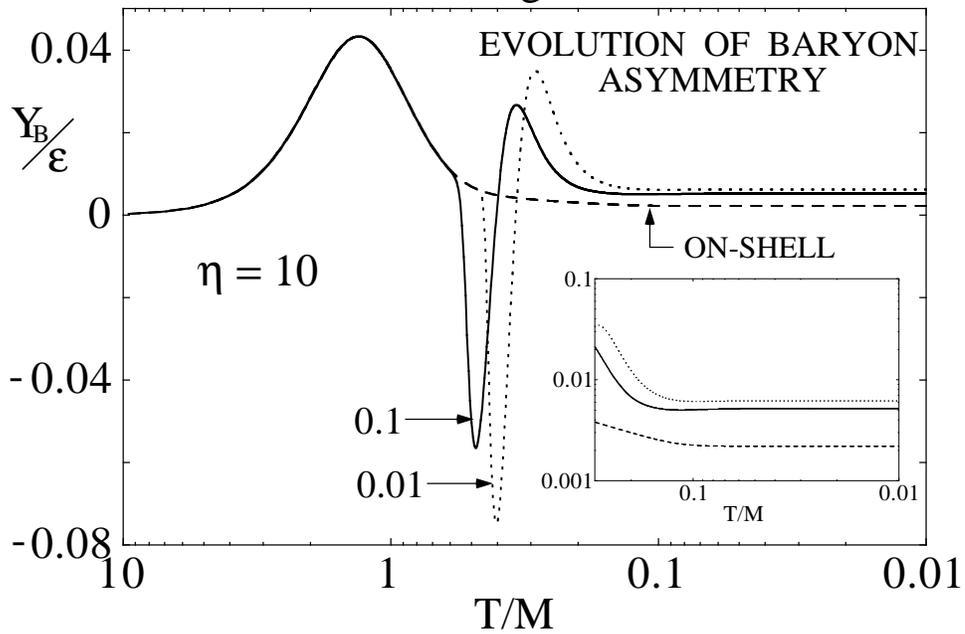
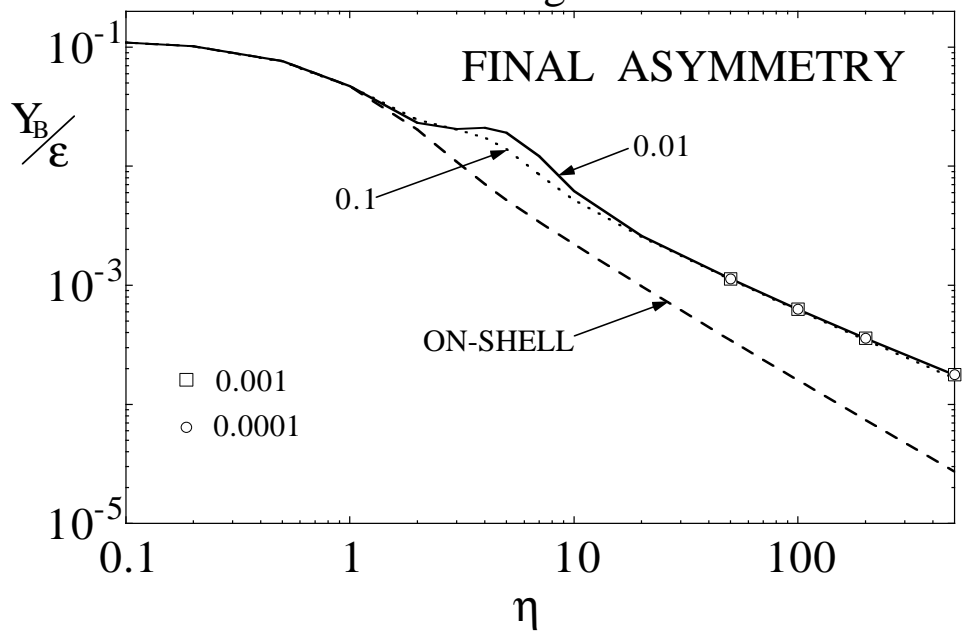
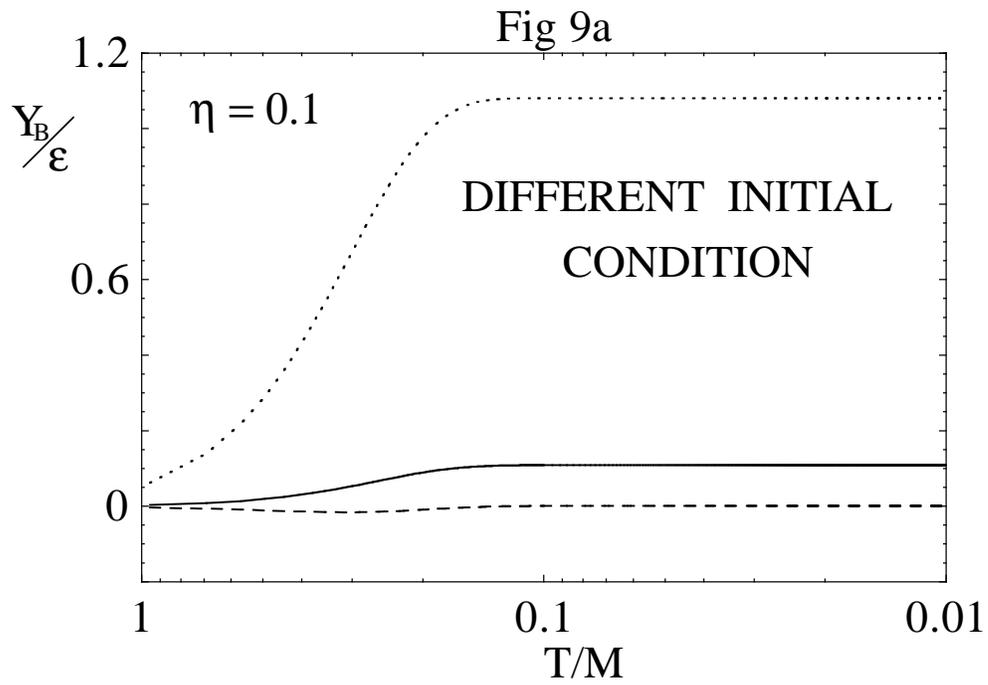
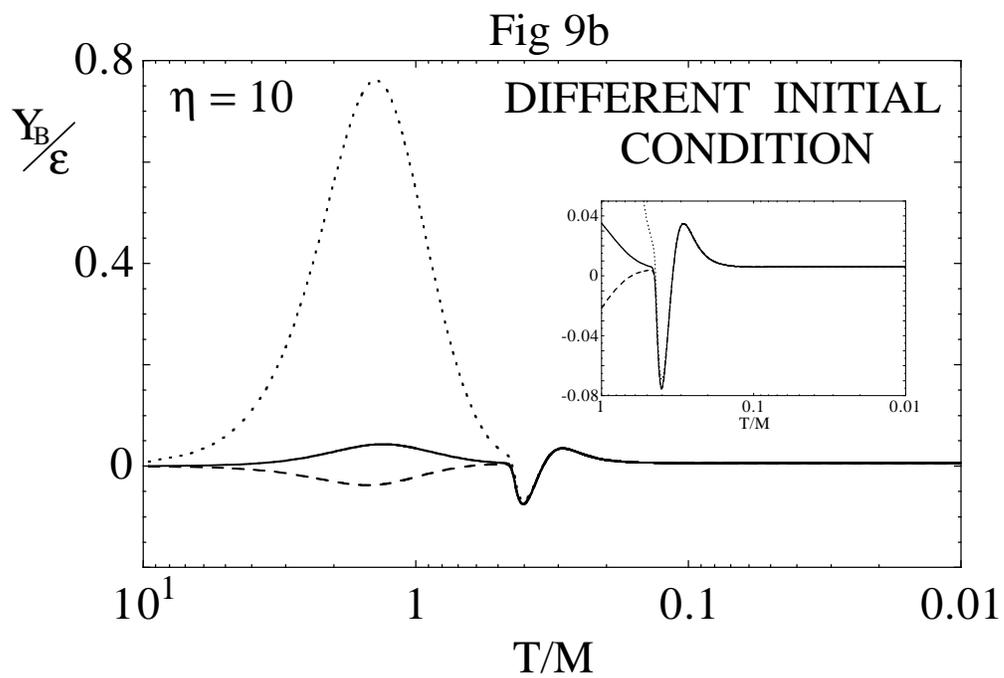
\end{document}